\documentclass[11pt,a4paper]{article}
\pdfoutput=1
\usepackage{jheppub}
\usepackage{amsmath,amsfonts,epsfig,graphics}
\usepackage{amssymb}
\usepackage{mathrsfs} 
\usepackage{amsbsy,color}
\usepackage{latexsym}
\usepackage{graphicx}
\usepackage{psfrag}
\usepackage{multirow}
\usepackage{booktabs}
\usepackage{url}
\usepackage{dcolumn}
\usepackage[T1]{fontenc}

\newcommand{\be}{\begin{equation}}
\newcommand{\ee}{\end{equation}}
\newcommand{\bea}{\begin{eqnarray}}
\newcommand{\eea}{\end{eqnarray}}
\newcommand{\bes}{\begin{eqnarray}}
\newcommand{\ees}{\end{eqnarray}}
\newcommand{\ba}{\begin{array}}
\newcommand{\ea}{\end{array}}

\newcommand{\Eq}[1]{Eq.~(\ref{#1})}

\newcommand{\Eqs}[1]{Eqs.~(\ref{#1})}
\newcommand{\fig}[1]{Fig.~\ref{#1}}
\newcommand{\Fig}[1]{Figure~\ref{#1}}

\newcommand{\Sect}[1]{Section~\ref{#1}}

\newcommand{\Tab}[1]{Table~\ref{#1}}

\newcommand{\Ref}[1]{Ref.~\cite{#1}}

\newcommand{\tauint}{\tau_\mathrm{int}}
\newcommand{\tauexp}{\tau_\mathrm{exp}}

\newcommand{\SW}{S_\mathrm{W}}

\newcommand{\nf}{N_\mathrm{f}}

\newcommand{\rmO}{\mathrm{O}}
\newcommand{\csw}{c_\mathrm{SW}}
\newcommand{\kud}{\kappa_\mathrm{ud}}
\newcommand{\mud}{m_\mathrm{ud}}
\newcommand{\ms}{m_\mathrm{s}}
\newcommand{\Qtop}{Q_\mathrm{top}}
\newcommand{\ks}{\kappa_\mathrm{s}}

\newcommand{\fm}{\mathrm{fm}}
\newcommand{\mev}{\mathrm{MeV}}

\pdfoutput=1   

\begin{document}

\title{ Simulation of QCD with $N_\mathrm{f}=2+1$ flavors of non-perturbatively
improved Wilson fermions }

\author[a]{Mattia~Bruno}
\author[b]{Dalibor~Djukanovic}
\author[c]{Georg~P.~Engel}
\author[b]{Anthony~Francis}
\author[d]{Gregorio~Herdoiza}
\author[e]{Hanno~Horch}
\author[a]{Piotr~Korcyl}
\author[f]{Tomasz~Korzec}
\author[g]{Mauro~Papinutto}
\author[a]{Stefan~Schaefer}
\author[h]{Enno~E.~Scholz}
\author[h]{Jakob~Simeth}
\author[a]{Hubert~Simma}
\author[h]{Wolfgang~S\"oldner}

%
%
\affiliation[a]{John von Neumann Institute for Computing (NIC), DESY \\
    Platanenallee 6, D-15738 Zeuthen, Germany}
\affiliation[b]{Helmholtz Institute Mainz, University of Mainz, D-55099 Mainz, Germany}
\affiliation[c]{Dipartimento di Fisica, Universit\`a di Milano-Bicocca, and
INFN, Sezione di Milano-Bicocca\\ Piazza della Scienza 3,
I-20126 Milano, Italy
}
\affiliation[d]{Departamento de F\'isica Te\'orica and Instituto de F\'isica Te\'orica
    UAM/CSIC,\\ Universidad Aut\'onoma de Madrid, Cantoblanco, E-28049
        Madrid, Spain}
\affiliation[e]{PRISMA Cluster of Excellence, Institut f\"ur Kernphysik, \\Johannes Gutenberg Universit\"at Mainz, D-55099 Mainz, Germany}
\affiliation[f]{Institut f\"ur Physik, Humboldt Universit\"at \\
  Newtonstr.~15, D-12489 Berlin, Germany}
\affiliation[g]{Dipartimento di Fisica, ``Sapienza'' Universit\`a di Roma, and
INFN, Sezione di Roma,\\ P.le Aldo Moro 2,
I-00185 Roma, Italy}
\affiliation[h]{Institut f\"ur Theoretische Physik,
Universit\"at Regensburg, \\D-93040 Regensburg, Germany}

\abstract{
We describe a new set of gauge configurations generated within
the CLS effort. These ensembles have $N_\mathrm{f}=2+1$ flavors of non-perturbatively
improved Wilson fermions in the sea with the L\"uscher--Weisz action used for
the gluons. Open boundary conditions in time are used to address the problem of
topological freezing at small lattice spacings and twisted-mass reweighting for
improved stability of the simulations. We give the bare parameters at which the
ensembles have been generated and how these parameters have been chosen.
Details of the algorithmic setup and its performance are presented as well as
measurements of the pion and kaon masses alongside the scale parameter $t_0$.
}


\maketitle

\section{Introduction}

What can be achieved in lattice QCD computations in terms of observables 
and their accuracy depends to a large extent on the availability of
suitable ensembles of gauge field configurations.  
For reliable results, many sources of error have to be controlled:
Fine lattices are needed for minimal discretization effects, 
the quark masses have to be close to their physical values and the 
volume of the lattices has to be large enough for finite size effects
to be small. The final precision also depends on the quark flavor 
content of the sea. On top of these systematic effects come statistical
uncertainties: Simulations have to be long enough such  that the 
statistical errors can be estimated reliably, and with statistical uncertainties
getting smaller, the need for control over systematic effects increases.

Since the generation of gauge field configurations is computationally the most
demanding part of the whole computation, a careful evaluation of the physics
parameters needs to be made  in view of the target precision of the observables
--- as far as this is possible at this stage. The goal is to balance the
various sources of systematic and statistical uncertainties in the final result:
In light of the findings of Ref.~\cite{Bruno:2014ufa}, for example, we do not include a
dynamical charm quark in the sea as we do not anticipate to be able to reach an
accuracy comparable to its effect on typical low-energy observables after taking
the continuum limit and the chiral extrapolation. 
On the contrary, including the charm might introduce large
lattice artifacts and would make the tuning procedure more difficult.

Recent year's advances have led to a re-evaluation of the requirements for a
reliable lattice computation regarding the control over statistical errors.
Notably, it has been known for a while that the global topological charge
freezes on fine lattices with periodic boundary
conditions~\cite{DelDebbio:2002xa,Bernard:2003gq,Schaefer:2010hu}.  However,
with the advent of the gradient flow in lattice
computations~\cite{Luscher:2010we,Luscher:2010iy} it has been discovered that
at moderate lattice spacing other quantities constructed from smoothed fields
evolve even slower in Monte Carlo time~\cite{Luscher:2011kk,Bruno:2014ova}. To
exclude uncontrolled biases  in any observable, Monte Carlo histories much
longer than the exponential autocorrelation time are required, i.e. much
longer than the times observed in these smoothed observables and therefore longer
than previously thought.

In this paper, we give an overview of the first round of the CLS (Coordinated
Lattice  Simulations) effort to generate configurations with $\nf=2+1$ flavors
of non-perturbatively improved Wilson fermions. In some of its
aspects it is a continuation of the $\nf=2$ flavor project: We use a 
non-perturbatively
improved  Wilson fermion action, we do not employ link smearing, simulations are
done using a public code and we focus on small lattice spacings for a controlled continuum
limit~\cite{Fritzsch:2012wq}.

By adding the additional flavor to the sea, one naturally aims at higher
accuracy than with two flavor simulations. In order to achieve this, there are
also improvements over the previous project. We use open boundary conditions in
the time direction which prevent the topological charge from
freezing~\cite{Luscher:2010we,Luscher:2011kk} and twisted-mass reweighting to
avoid the sector formation due to zero eigenvalues of the Wilson
fermions and the resulting instabilities in the
simulation~\cite{Luscher:2008tw}. 

The simulations are  performed using the openQCD code version
1.2~\cite{openQCD} whose general algorithmic setup is described in
\Ref{Luscher:2012av}. The code provides broad flexibility with respect to the
algorithms used, starting from the determinant decomposition, the molecular
dynamics integration and the methods employed for solving the Dirac equation.
In this paper we give the physics and algorithmic parameters of these
simulations and report on our experiences with this new setup. Furthermore we
present first measurements of basic physics observables: the masses of the pion
and the kaon as well as the scale parameter $t_0$~\cite{Luscher:2010iy} on
which we base the tuning of the runs.

Similar large-scale simulations of QCD have recently been performed by  the
PACS-CS simulating improved Wilson fermions~\cite{Aoki:2009ix}, the  QCDSF
collaboration with $\nf=2+1$ flavors of NP improved, smeared Wilson
fermions~\cite{Bietenholz:2010jr}, the Hadron Spectrum collaboration using
$\nf=2+1$ flavors of tree-level improved, smeared Wilson fermions on
anisotropic lattices~\cite{Lin:2008pr}, as well as the ETM collaboration using
twisted-mass fermions~\cite{Baron:2010bv} and the BMW collaboration with
tree-level improved smeared fermions~\cite{Borsanyi:2014jba}. Also domain wall
fermions are employed by RBC-UKQCD~\cite{Arthur:2012opa} and overlap fermions
by JLQCD~\cite{Aoki:2008tq} as well as smeared rooted staggered fermions by
MILC~\cite{Bazavov:2012xda}. Our simulations are unique by their use of  open
boundary conditions and, among the simulations with standard Wilson fermions,
twisted-mass reweighting as a safeguard against the effects  of near-zero modes
of the Dirac operator.

The paper is organized as follows: In \Sect{sec:phys} we give the details of
the action, the tuning strategy and the parameters of the runs. The algorithmic
setup is described in \Sect{sec:algo}. Autocorrelations observed in the
simulations are the subject of \Sect{sec:ac}, while the two types of
reweighting used in the light and the strange quark sector are discussed
in \Sect{sec:rw}. This is followed in \Sect{sec:meas} by the measurement of the
pseudoscalar masses and the scale parameter $t_0$ and a discussion of discretization effects
in \Sect{sec:beta33}.

\section{Physical parameters\label{sec:phys}}
The simulations are done on lattices of size $N_\mathrm{t}\times
N_\mathrm{s}^3$, with open boundary conditions imposed on time slice $0$ and
$N_\mathrm{t}-1$. Lattices with $N_\mathrm{t}$ points in the temporal
direction therefore have a physical time extent  of
$T=(N_\mathrm{t}-1)a$ in conventional notation, with $a$ being the lattice spacing. 

\subsection{Action}
The general setup of the lattice actions which can be simulated with the
openQCD code has already been given in detail in \Ref{Luscher:2012av}. In
particular it is described there how the boundary conditions are imposed. Therefore, 
here we only give  details of the bulk action. Throughout, the coefficients of the
boundary improvement terms are set to their tree level values.

For the gauge fields, we use the  L\"uscher--Weisz action~\cite{Luscher:1984xn}
with tree level coefficients --- which is different from the earlier reference~\cite{Luscher:2012av} where the Iwasaki
action has been employed. In the bulk, the plaquette and rectangle terms are
multiplied by their respective coefficients  $c_0=5/3$ and $c_1=-1/12$ 
\begin{equation} 
S_g[U]=\frac{\beta}{6} \big ( c_0\, \sum_{p} 
\mathrm{tr} \{ 1-U(p) \} + c_1 \sum_{r} \mathrm{tr} \{ 1-U(r) \} \big )
\,, 
\label{eq:gauge}
\end{equation}
where the sums run over the plaquettes $p$ and the rectangles $r$ contained in
the lattice and $\beta=6/g_0^2$ with the bare gauge coupling $g_0$. 

For the fermions, the  Wilson Dirac operator~\cite{Wilson} including the Sheikholeslami--Wohlert term needed for
$\rm{O}(a)$ improvement of the action~\cite{impr:SW} is used
\begin{equation} D_\text{W}(m_0) = 
\frac{1}{2}\sum_{\mu=0}^3 \{\gamma_\mu(\nabla^*_\mu+\nabla_\mu) -a\nabla^*_\mu\nabla_\mu\}
+a\csw \sum_{\mu,\nu=0}^3\frac{i}{4}\sigma_{\mu\nu}\widehat F_{\mu\nu}+m_0
\end{equation}
with $\nabla_\mu$ and $\nabla_\mu^*$ the covariant
forward and backward derivatives, respectively. The improvement term
containing the standard discretization of the field strength
tensor $\widehat F_{\mu\nu}$~\cite{impr:pap1} comes with the coefficient $\csw$
whose value has been determined non-perturbatively in \Ref{Bulava:2013cta}.

The three flavor fermion action  then reads
\begin{equation}
S_\mathrm{f}[U,\overline \psi,\psi]=
a^4 \sum_{f=1}^3 \sum_x\, \overline \psi_f(x)\, D_\mathrm{W}(m_{0,f})\, \psi_f(x)\,,
\label{eq:ferm}
\end{equation}
where we take the up and down quark masses to be degenerate 
$m_{0,\mathrm{ud}}\equiv m_{0,\mathrm{u}}=m_{0,\mathrm{d}}$.
The strange-quark mass $m_{0,\mathrm{s}}$ is tuned as a function of the
light quark mass.
In the following, we frequently quote the hopping parameters
$\kappa_\mathrm{f}$ instead of the bare
quark masses 
\begin{equation}
 m_{0,f}=\frac{1}{2a} (\frac{1}{\kappa_\mathrm{f}}-8)\,.
\end{equation}

\subsection{Choice of parameters}
Since we do not simulate the full Standard Model but restrict ourselves to $\nf=2+1$ 
flavor QCD, electromagnetic and isospin breaking effects as well as the contributions from 
the heavy sea quarks, among others, are not included in this calculation.  Therefore
the point of ``physical'' quark masses is not unique even in the continuum and we
have to fix observables which define it. For the tuning of our runs, we set the scale 
through $t_0$ defined by the Wilson flow~\cite{Luscher:2010iy}, see \Sect{sec:t0}. 
The quark masses are set using the masses of the pion and the kaon. While this choice is 
convenient during the tuning of the runs, it can be changed in the future once more observables are
available.

The lattices at different cutoff are matched via the dimensionless parameters
\begin{align}
\phi_2&=8 t_0 m_\pi^2\, &&\text{and} & \phi_4&=8 t_0 (m_K^2+\frac{1}{2}m_\pi^2)\,,
\label{eq:p42}
\end{align}
where all quantities are the ones measured at the parameter values of the
ensemble in question. Note that in leading order of Chiral Perturbation Theory (ChPT)
they are proportional to the sum of the quark masses, $\phi_2\propto
(m_\mathrm{u}+m_\mathrm{d})$ and $\phi_4\propto
(m_\mathrm{u}+m_\mathrm{d}+m_\mathrm{s})$~\cite{GellMann:1968rz,Bar:2013ora}.
The advantage of this strategy is that we obtain all quantities involved with
high statistical accuracy from the simulated ensembles, without further need of
renormalization constants or chiral extrapolation.

Particular drawbacks of this strategy are the significant cutoff
effects which we observe in the various definitions of $t_0/a^2$ on our largest
lattice spacings, as discussed in \Sect{sec:t0cut}. Furthermore, the value of
$t_0$ is not an experimentally accessible observable  and only known from other
lattice simulations. In the literature one finds
$\sqrt{8t_0}=0.4341(33)\,$fm  by the ALPHA collaboration
using Wilson fermions  in two-flavor QCD~\cite{Bruno:2013gha} and $\sqrt{8t_0}=0.4144(59)(37)\,$fm
by the BMW collaboration using $\nf=$2+1 flavors~\cite{Borsanyi:2012zs}. In a $2+1+1$
flavor setup with rooted staggered fermions, the HPQCD collaboration finds
$\sqrt{8 t_0}=0.4016(23)\,$fm~\cite{Dowdall:2013rya}.
As has been observed in \Ref{Bruno:2013gha}, these numbers exhibit a
significant flavor content effect, which however is monotonic in the number of
flavors. Since our simulation setup is also with $\nf= 2+1$ flavors, we choose 
the value of \Ref{Borsanyi:2012zs}. 

The QCD values of  $m_\pi=134.8(3)$~MeV
and $m_K=494.2(4)$~MeV in the isospin limit and without electromagnetic contributions
are taken from the analysis of \Ref{Aoki:2013ldr}. The correction of the experimental
masses is based on ChPT at NLO with input from other lattice calculations showing a 
suppression of the contribution from the combination of low-energy constants relevant to this
case.
 This leads to a physical point estimate
\begin{align}
\phi_2^\mathrm{phys}&=0.0801(27)\,, & \phi_4^\mathrm{phys}&=1.117(38),
\end{align}
where errors have been added in quadrature.

From this choice and our measurements of $t_0/a^2$ presented below, we estimate for 
our three values of  $\beta=3.4$, $3.55$ and $3.7$ 
lattice spacings $a$ of $a\approx 0.086\,\fm$, $0.064\,\fm$ and $0.05\,\fm$, respectively.

\subsection{Quark mass trajectory}

In order to achieve $\rmO(a)$ improvement, the bare coupling --- as all
bare parameters --- has to be improved with a mass-dependent 
term~\cite{impr:pap1}
\begin{equation}
\tilde g_0^2 = g_0^2 \,\big\{1+\frac{b_\mathrm{g}}{3} a\sum_f(m_{0,\mathrm{f}}-m_\mathrm{cr}) \big\}\,,
\end{equation}
with  $m_\mathrm{cr}$ the critical quark mass whose precise value is not
known at this stage. To keep the lattice spacing constant as we change
the sea quark masses, this modified coupling constant $\tilde g_0$ has to be kept constant.

While the coefficient $b_\mathrm{g}$ is small at one loop in perturbation
theory~\cite{impr:pap5}, $b_\mathrm{g}=0.012\,\nf\, g_0^2$, a
non-perturbative result is not known for any action. To keep $\tilde g_0$
fixed, we therefore keep the sum over the subtracted quark masses fixed, 
a strategy already proposed in
\Ref{Bietenholz:2010jr}. Note that this is equivalent to keeping  the sum
over the bare quark masses $m_{0,\mathrm{f}}$ fixed
\begin{align}
a\sum_{f=1}^3 (m_{0,\mathrm{f}}-m_\mathrm{cr}) &= \text{const} & \Leftrightarrow &&
a\sum_{f=1}^3 m_{0,\mathrm{f}} &= \text{const} 
& \Leftrightarrow &&  \sum_{f=1}^3 \frac{1}{\kappa_\mathrm{f}}  = \text{const} \,.
\label{eq:chitra}
\end{align}
Up to effects of order $\rmO(a\mud)$, this also implies a constant sum of
improved PCAC quark masses \cite{impr:nondeg}.

We can therefore define chiral trajectories by a point in the $\phi_2$--$\phi_4$ plane,
at which different lattice spacings are matched and the requirement that the sum
of the bare quark masses is constant. For each value of $\beta$, the lattice spacing is
constant along these lines and a continuum limit can be performed for each value of $\phi_2$. 

As explained in the next section, 
we match lattices with different lattice spacings at $m_\pi=m_\mathrm{K}\approx 420\,\mev$,
where we will also show first results concerning the size of the $\mathrm{O}(am)$ 
cutoff effects introduced by this choice.

\subsection{Tuning strategy}
By choosing the chiral trajectories of \Eq{eq:chitra}, the tuning process
can be highly simplified: Keeping $\beta$ fixed, for each chiral trajectory
we  match the lattices at the flavor symmetric point, i.e. where
all quarks have equal masses.

Determining the slope of $\phi_4$ as a
function of $\phi_2$ at $\beta=3.4$  from a set of preliminary runs, not shown here,
we estimate the target value on the symmetric line
\begin{equation}
\phi_4 \big |_{\mud=\ms}=1.15 \,.
\end{equation}
With the final statistics, we are able to reach better than 1\% accuracy in this quantity
and a matching of the target value within one standard deviation. In
the chiral limit this translates into an accuracy of about $1\,\mev$ in the
strange-quark mass. 
In the future, we
plan to have more chiral trajectories which will allow us to study the
consequences of the remaining mistuning.

\begin{figure}
\begin{center}
\includegraphics[width=0.7\textwidth]{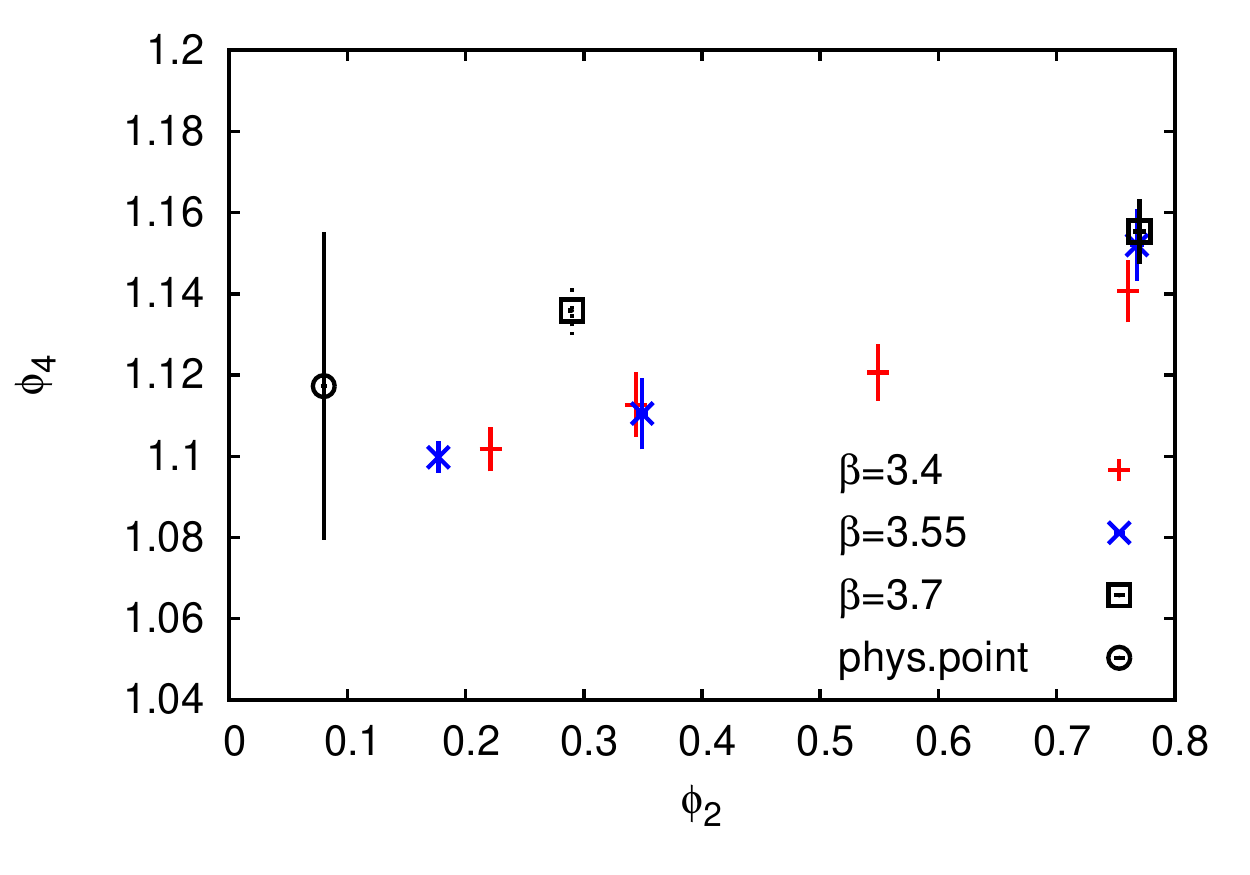}
\end{center}
\caption{Position of our ensembles in terms of the dimensionless variables 
$\phi_2$ and $\phi_4$ defined in \Eq{eq:p42}. The rightmost points are on the symmetric 
line $m_\mathrm{ud}=m_\mathrm{s}$. The fact that the point with the smallest $\phi_2$
at $\beta=3.7$ is above the range indicated by coarser lattices might be an effect of mistuning
at the symmetric point or an indication for underestimated errors due to the low statistics
indicated by the dashed error bars.\label{fig:p42}}
\end{figure}

The result of the tuning effort and the resulting trajectory in the
$\phi_2$--$\phi_4$ plane is shown in \Fig{fig:p42} 
with results from the ensembles  given in \Tab{tab:ens}. Within the statistical
accuracy, we do not observe significant cutoff effects. The one point at $\beta=3.7$
is still under production and its error therefore not yet trustworthy. We observe, the
quark mass effect on $\phi_4$ along this trajectory is moderate, around $5\%$
between the chiral limit and the symmetric point, as expected from ChPT.

\begin{table}[tb]
\begin{center}
\small
\begin{tabular}{ccccllcccc}
\toprule
id &   $\beta$ &  $N_\mathrm{s}$  &  $N_\mathrm{t}$  &  $\kappa_u$ & $\kappa_s$ & $m_\pi$[MeV] &   $m_K$[MeV] &  $m_\pi L$\\
 \midrule
B105 & 3.40 & 32 & 64	& 0.136970 & 0.13634079		& 280 &460  & 3.9\\
H101 & 3.40 & 32 & 96	& 0.13675962 & 0.13675962	& 420 &420  & 5.8\\
H102 & 3.40 & 32 & 96	& 0.136865 & 0.136549339	& 350 &440  & 4.9\\
H105 & 3.40 & 32 & 96	& 0.136970 & 0.13634079		& 280 &460  & 3.9\\
C101 & 3.40 & 48 & 96	& 0.137030 & 0.136222041	& 220 &470  & 4.7 \\
D100 & 3.40 & 64 & 128	& 0.137090 & 0.136103607	& 130 &480  & 3.7 \\
\midrule                                                   
H200 & 3.55 & 32 & 96	& 0.137000 & 0.137000   & 420	&420  & 4.4 	\\
N200 & 3.55 & 48 & 128	& 0.137140 & 0.13672086	& 280&  460  & 4.4 	\\
D200 & 3.55 & 64 & 128	& 0.137200 & 0.136601748& 200 &	480  & 4.2		\\
\midrule		                                               
N300 & 3.70 & 48 & 128	& 0.137000 & 0.137000	 & 420	&420  & 5.1 	\\
N301 & 3.70 & 48 & 128	& 0.137005 & 0.137005	 & 410	&410  & 4.9 	\\
J303 & 3.70 & 64 & 192	& 0.137123 & 0.1367546608& 260	&470  & 4.1		\\
\bottomrule
\end{tabular}
\caption{\label{tab:ens}List of the ensembles. In the id, the letter gives the geometry, the 
first digit the coupling and the final two label the quark mass combination. We give rounded values
of $m_\pi$ and $m_\mathrm{K}$ using the $t_0/a^2$ of the ensemble and $\sqrt{8t_0}=0.4144\,\fm$.
Using $t_0/a^2$ extrapolated to the physical light quark masses, we estimate  lattice spacings 
of $a\approx 0.086\,\fm$, $0.064\,\fm$ and $0.05\,\fm$ for $\beta=3.4$, $3.55$ and $3.7$, respectively.}
\end{center} 
\end{table}

\section{Algorithmic parameters\label{sec:algo}}
The basic algorithmic setup has already been described in detail in
\Ref{Luscher:2012av}, but since we are presenting simulations with larger
lattices, statistics and a different action, the various settings needed to be
reconsidered. Here we give the parameters at which the runs were performed
and the reasoning behind the various choices.

\subsection{Twisted-mass reweighting\label{sec:tm}}
Since the Wilson Dirac operator is not protected against eigenvalues below
the quark mass, field space is  divided by surfaces of zero eigenvalues. These
barriers of infinite action cannot be crossed during 
the molecular dynamics evolution, at least if the equations of 
motion are integrated exactly.

While at a sufficiently large volume and quark mass this might not be a problem
in practice~\cite{DelDebbio:2005qa}, it can lead to instabilities during the
simulation and meta-stabilities in the thermalization phase.  L\"uscher and
Palombi~\cite{Luscher:2008tw} therefore suggested to introduce a small
twisted-mass term into the action during the simulation and compensate for this
by reweighting.

In the present simulations, we use the second version of the reweighting
suggested in \Ref{Luscher:2008tw}, which is less affected by fluctuations in
the reweighting factor from the ultraviolet part of the spectrum of the Dirac
operator. Contrary to the original proposal, we do not apply it to the
Hermitian Dirac operator $Q=\gamma_5 D_W$ but to the Schur complement $\hat
Q=Q_\mathrm{ee}-Q_\mathrm{eo} Q_\mathrm{oo}^{-1} Q_\mathrm{oe}$ of the
asymmetric even-odd preconditioning~\cite{DeGrand:1988vx}. This amounts to
replacing the determinant of the light quark pair by
\begin{equation}
\det Q^2= {\det}^2 Q_\mathrm{oo}\, \det \hat Q^2 
\to {\det}^2 Q_\mathrm{oo} \,
\det \frac{\hat Q^2+ \mu_0^2}{\hat Q^2 + 2\mu_0^2} 
\,
\det \big ( \hat Q^2 +\mu_0^2 \big ) \,.
\label{eq:detud}
\end{equation}
The reweighting factor which needs to be included in the measurement
of primary observables then reads
\begin{equation}
W_0=
\det \frac{(\hat Q^2+ 2\mu_0^2) \, \hat Q^2}{(\hat Q^2 + \mu_0^2)^2} \,.
\label{eq:w0}
\end{equation}
The choice of the parameter $\mu_0$ will be discussed in \Sect{s:tmrw}.

\subsection{Determinant factorization}
The fluctuations in the forces have to be reduced further than what can be
achieved by introducing an infrared cutoff by the twisted mass $\mu_0$. To this 
end we use  Hasenbusch's mass
factorization~\cite{algo:GHMC} with a twisted mass~\cite{Hasenbusch:2002ai}
applied to the last term in \Eq{eq:detud}~\cite{Schaefer:2012tq} 
\begin{equation}
\det \big ( \hat Q^2 +\mu_0^2 \big ) 
=  
\det \big (\hat Q^2 + \mu_{N_\mathrm{mf}}^2 \big )
\times 
\prod_{i=1}^{N_\mathrm{mf}} 
\det \frac{\hat Q^2 + \mu_{i-1}^2 }{\hat Q^2 + \mu_{i}^2 } 
\label{eq:mf}
\end{equation}
with a tower of increasing values of $\mu_0<\mu_1<\dots<\mu_{N_\mathrm{mf}}$.
The values of these masses can significantly influence the performance of the
algorithm. Here we roughly set them at equal distances on a logarithmic
scale as suggested in \Ref{Luscher:2012av}. The precise values of the 
$\mu_i$ are listed in \Tab{tab:algo}, which implicitly gives also the number of 
factors in \Eq{eq:mf}.

\setlength{\tabcolsep}{1pt}
\begin{table}
\begin{center}
\small
\begin{tabular}{ccccccccccc}
\toprule
id 	&$a\mu_0$	&$a\mu_i$		&$N_{\mathrm{mf},2}$	&$N_\mathrm{p}$	&$[r_a,r_b]$	&$N_{\mathrm{p}}'$	&$N_{\mathrm{p},2}$	&$N_{\mathrm{s},2}$	&MDU 	&$\langle P_\mathrm{acc}\rangle$\\
\midrule
B105r002& $0.001$ & $\{0.005,0.05,0.5\}$	&1	&10&$[0.0200,7.00]$ &3	&2					&8					&1984	&0.99 \\
B105r003&$0.001$&$\{0.005,0.05,0.5\}$		&1	&10&$[0.0170,7.80]$ &3	&2					&8					&4120	&0.96 \\
H101r000 &$0.001$&$\{0.005,0.05,0.5\}$		&2	&12&$[0.0056,7.50]$ &5	&2					&10					&4028	&0.95 \\
H101r001 &$0.001$&$\{0.005,0.05,0.5\}$		&2	&12&$[0.0056,7.50]$ &5	&2					&10					&4036	&0.95 \\
H102r001 &$0.001$&$\{0.005,0.05,0.5\}$		&1	&12&$[0.0070,7.40]$ &6	&4					&10					&4116	&0.97 \\
H102r002 &$0.001$&$\{0.005,0.05,0.5\}$		&1	&12&$[0.0080,7.60]$ &6	&4					&10					&4032	&0.97 \\
H105r001 &$0.001$&$\{0.005,0.05,0.5\}$		&1	&11&$[0.0100,7.30]$ &4	&2					&10					&4108	&0.97 \\
H105r002 &$0.001$&$\{0.005,0.05,0.5\}$		&1	&11&$[0.0100,7.30]$ &4	&2					&10					&4168	&0.98 \\
H105r005 &$0.0005$&$\{0.005,0.05,0.5\}$		&1	&13&$[0.0032,7.60]$ &6	&3					&7					&3348	&0.89 \\ 
C101r010 &$0.0006$&$\{0.007,0.05,0.5\}$		&1	&12&$[0.0085,7.80]$ &5	&2					&9					&1404	&0.84 \\
C101r013 &$0.0003$&$\{0.007,0.05,0.5\}$		&1	&13&$[0.0060,7.80]$ &6	&3					&13					&~~868	&0.95 \\
C101r014 &$0.0006$&$\{0.007,0.05,0.5\}$		&1	&13&$[0.0060,7.80]$ &6	&3					&12					&2100	&0.95 \\
C101r015 &$0.0003$&$\{0.007,0.05,0.5\}$		&1	&13&$[0.0060,7.80]$ &6	&3					&13					&2402	&0.90 \\
D100r002&$0.0001$&$\{0.00016,0.0005,$ 		&1	&14&$[0.0030,8.15]$ &7	&2					&18					&~~178	&0.69 \\ 
	&	&$0.0055,0.06,0.7\}$	\\
\midrule
H200r000 &$0.001$&$\{0.005,0.05,0.5\}$		&1	&12&$[0.0050,6.50]$ &6	&3					&10					&4000	&1.00 \\ 
H200r001 &$0.001$&$\{0.005,0.05,0.5\}$		&1	&12&$[0.0050,6.50]$ &6	&3					&10					&4000	&1.00 \\ 
N200r000&$0.00065$&$\{0.005,0.05,0.5\}$		&1	&12&$[0.0100,7.10]$ &6	&3					&7					&3424	&0.94 \\ 
N200r001&$0.00065$&$\{0.005,0.05,0.5\}$		&1	&12&$[0.0100,7.10]$ &6	&3					&7					&3424	&0.94 \\ 
D200r000&$0.0003$&$\{0.00075,0.005,$		&1	&13&$[0.0060,7.80]$ &6	&2					&8					&3572	&0.94 \\ 
	&	&$0.05,0.5\}$	\\		
\midrule
N300r002 &$0.001$&$\{0.01,0.05,0.5\}$		&1	&13&$[0.0050,7.20]$ &6	&3					&6					&6162	&0.94 \\ 
N301r000 &$0.001$&$\{0.01,0.05,0.5\}$		&1	&13&$[0.0050,6.00]$ &6	&3					&6					&1944	&0.95 \\
N301r001 &$0.001$&$\{0.01,0.05,0.5\}$		&1	&13&$[0.0050,6.00]$ &6	&3					&6					&1852	&0.95 \\
J303r003 &$0.00075$&$\{0.002625,0.009188,$	&1	&13&$[0.0080,7.00]$ &7	&3					&6					&2328	&0.88 \\ 
	&	&$0.032156,0.112547,0.5\}$	\\
\bottomrule
\end{tabular}
\caption{\label{tab:algo}Parameters of the algorithm: We give the twisted masses used 
in the reweighting and mass factorization, the $N_{\mathrm{mf},2}$
lightest of which are integrated on the coarsest time scale, 
the number of poles $N_\mathrm{p}$ and the range used in the RHMC, with $N_{\mathrm{p}}'$ put
on single pseudofermions ,  $N_{\mathrm{p},2}$ of which  are
integrated on the outer level. $N_{\mathrm{s},2}$ is the number of steps of the outer level
of the MD integrator used for one trajectory. The total length of the Markov chain and the acceptance rate are
also given.
}
\end{center}
\end{table}

The combination of twisted-mass reweighting and mass factorization leads to an 
effective action for the light quark pair with $N_\mathrm{mf}+2$ terms
\begin{equation}
\begin{split}
S_\mathrm{ud, eff}[U,\phi_0,\dots,\phi_{N_\mathrm{mf}+1}]=&
\big (\phi_0, \frac{\hat Q^2+ 2\mu_0^2}{\hat Q^2 + \mu_0^2}  \phi_0 \big)
+
\sum_{i=1}^{N_\mathrm{mf}} 
\big(\phi_i, \frac{\hat Q^2 + \mu_{i}^2 }{\hat Q^2 + \mu_{i-1}^2 }  \phi_i \big )
\\
&
+ \big \{ \big (\phi_{N_\mathrm{mf}+1}, \frac{1}{\hat Q^2 + \mu_{N_\mathrm{mf}}^2} \phi_{N_\mathrm{mf}+1} \big )
-2 \log \det  Q_\mathrm{oo} \} \,.
\label{eq:effmf}
\end{split}
\end{equation}
The single term with the largest twisted mass and  the one from the diagonal
determinant $\det  Q_\mathrm{oo}$ are always integrated together and are therefore
counted as one term.

\subsection{RHMC}
The strange quark is simulated using the RHMC
algorithm~\cite{Kennedy:1998cu,Clark:2006fx}, where the matrix square root is
approximated by a rational function
\begin{equation}
\det Q = \det Q_\mathrm{oo} \, \det \sqrt{ \hat Q^2 } =  \det Q_\mathrm{oo} \, 
{\det} \big (A^{-1} \prod_{i=1}^{N_\mathrm{p}} \frac{\hat Q^2 +\bar \mu_{i}^2}{\hat Q^2 +\bar \nu_{i}^2} \big ) \times W_1\,.
\label{eq:rhmc}
\end{equation}
Zolotarev's optimal approximation to the inverse square root in the interval
$[r_a,r_b]$ with a given number of poles $N_\mathrm{p}$ determines the
parameters $A$ and  $\{\bar \mu_i,\bar \nu_i\}$. The strange-quark mass as argument of $Q$ and
$\hat Q$ has been suppressed for readability. $W_1$ is the reweighting factor,
implicitly defined by \Eq{eq:rhmc}, which has to be included in the
measurement. The values used in the various runs are specified in
\Tab{tab:algo}.

The openQCD code gives the option to split the determinant of the rational
function in \Eq{eq:rhmc} into several factors. In our simulations, we represent
the  $N_\mathrm{p}'$ terms with the smallest $\bar \mu_i$ of the product \Eq{eq:rhmc} by single 
pseudofermions, whereas the determinant of the remaining factors is expressed
as a single pseudofermion integral
\begin{equation}
\begin{split}
S_\mathrm{s, eff}[U,\phi_0,\dots,\phi_{N_\mathrm{p}'}]=&
\sum_{i=0}^{N_\mathrm{p}'-1}
\big(\phi_{i}, \frac{\hat Q^2 +\bar \nu_{N_\mathrm{p}-i}^2}{\hat Q^2 +\bar \mu_{N_{\mathrm{p}-i}}^2}
\phi_{i}\big)
+ \big (\phi_{N_\mathrm{p}'},  \prod_{j=1}^{N_\mathrm{p}-N_{\mathrm{p}}'} 
\frac{\hat Q^2 +\bar \nu_{j}^2}{\hat Q^2 +\bar \mu_{j}^2} \phi_{N_\mathrm{p}'} \big )
\\
&-\log \det Q_\mathrm{oo} \,,
\end{split}
\label{eq:effrhmc}
\end{equation}

Here again, the contribution from the two final terms is always considered
together. This decomposition has several advantages. First of all, the small
residues frequently  can be integrated on a larger time scale, due to a small
coefficient decreasing the forces. Furthermore, while the multi-shift conjugate
gradient algorithm~\cite{Jegerlehner:1996pm} is efficient for the combined
solution of the systems in the last factor with the large shifts, it turns out
to be advantageous to employ the deflated solver for the terms involving the
smaller $\bar \mu_i$. In this case it is no longer necessary to use a single
pseudofermion field for all shifts.

The range of the rational approximation is given by the smallest and the
largest eigenvalue of $\hat Q^2$ over typical gauge field configurations. On
thermalized configurations, estimates of these numbers can be obtained in
openQCD by the power method applied to $\hat Q^{-2}$ and  $\hat Q^{2}$,
respectively.  Typically, $\mathrm{O}(20)$ iterations proved sufficient for the
lower bound, whereas the largest eigenvalue required $\mathrm{O}(100)$
iterations. In particular the smallest eigenvalue turned out to be sensitive to
thermalization effects and exhibit larger fluctuations than expected. This made
it necessary to monitor it carefully at the beginning of each production run.

\subsection{HMC and the integration of the molecular dynamics}

In the algorithm the  action is split into  different components: the gauge action, the
determinants from the Hasenbusch splitting for the light quarks and the
various contributions to the strange-quark determinant from the rational
approximation described above, $N_\mathrm{mf}+N_\mathrm{p}'+4$ components in total. The
complete action  is simulated with the Hybrid Monte Carlo (HMC)
algorithm~\cite{Duane:1987de}; the  classical equations of motion are solved
numerically for trajectories of length $\tau=2$ in all simulations. This leads to Metropolis
proposals which are accepted with an acceptance rate $\langle P_\mathrm{acc}
\rangle$, given for our runs in \Tab{tab:algo}.

The goal of the splitting of the action, and the forces deriving
from it, is the reduction of the computational
cost needed to obtain a high acceptance rate at the end of the trajectory.  The
gauge forces are much cheaper to compute than the fermion forces, whose
components differ by orders of magnitude in size and fluctuations. It is therefore natural to
use a hierarchical integration scheme for the molecular dynamics of the HMC to
reflect this~\cite{Sexton:1992nu}. 

We use the setup described in \Ref{Luscher:2012av}, i.e., a three-level scheme
with the gauge fields integrated on the innermost level with the fourth order
integrator suggested by Omelyan, Mryglod, and Folk (OMF)~\cite{Omelyan2003272}
and implemented in the openQCD code. Most of the fermion forces are on the
intermediate level, again integrated with the fourth order integration scheme.
Only particularly small components of the fermion forces, that contribute
little to energy violation, are integrated on a larger scale with the
second order OMF integrator~\cite{Omelyan2003272}, whose parameter $\lambda$ is
set to $1/6$. 

Since one step of an inner level integration scheme is done for each outer step,
there are three parameters which define the scheme: the number of outermost steps
per trajectory $N_{\mathrm{s},2}$ and the number of poles 
$N_{\mathrm{p},2}$ as well as the number $N_{\mathrm{mf},2}$ of terms of \Eq{eq:mf} 
integrated on the outermost level. In the latter two cases the numbers refer to the 
terms with the smallest twisted-mass shifts. The values chosen in our runs 
can be found in \Tab{tab:algo}.

The choice of the trajectory length affects the autocorrelation times and is
therefore not easily studied. In general, longer trajectories have proven to be
beneficial~\cite{Schaefer:2010hu}, but in particular with dynamical fermions one
might prefer shorter trajectories because of instabilities of the integrator.
As a compromise, we use $\tau=2$. Asymptotically, this leads  to autocorrelations
growing with $\tauint \propto a^{-2}$. Note, however, that this scaling
behavior is also expected if the length of the trajectory is
scaled~\cite{Luscher:2011qa}.

\subsection{Solver}
The extensive use of the locally deflated
solver~\cite{Luscher:2007se,Luscher:2007es,Frommer:2013fsa} is an important
part of the progress that  made the presented simulations possible. It
removes the largest part of the cost increase as the quark mass is lowered,
thereby circumventing the significant slowing down observed in the past. The
increase in performance of the solver comes at the price of a  more complex
setup and many additional parameters which have to be chosen.

Fortunately, relatively little tuning of the local deflation subspace was
necessary here and we therefore do not list the parameters in detail. For most
runs, we used deflation blocks of size $4^4$. The parallelization of the $N_\mathrm{s}=48$ 
lattices required one or two dimensions to be set to $6$; also blocks
of $8\times 4^3$ have been used.

The number  of deflation modes per block has been chosen between $20$ and $32$,
in order  to balance the higher efficiency provided by the larger subspace and
the cost associated with the application of the preconditioner.

For the smaller lattices with $L/a=32$, we set the solver accuracy (the ratio
between the norm of the residue to the norm of the right hand side of the
equation) to $10^{-11}$ in the action and  $10^{-10}$ in the force computation.
To ensure the value of the action and the reversibility of the integration of
the equations of motion are sufficiently precise, more stringent residues have
been used for the lattices of larger volume.

\subsection{Production cost}
To give an idea of the cost of the various ensembles, we show in \Tab{tab:cost}
the average wall-clock time per molecular dynamics unit, along with the 
machine on which the run has been performed, the local lattice geometry, 
and the total number of cores used. 
Most of our production runs were carried out either on SuperMUC, a petascale 
cluster of IBM System x iDataPlex servers with Intel Sandy Bridge-EP processors
(Xeon E5-2680 8C) and Infiniband network (FDR10), or on IBM BlueGene/Q systems at 
CINECA (FERMI) and JSC (JUQUEEN). Since our code is not multi-threaded, we launch
four MPI processes per core on the BlueGene/Q machines to maximize overall performance.

Note that the execution times of the simulations do not only depend on the
algorithmic parameters (even for a single specific trajectory), but also on the
particular hardware on which the code has been run, as well as on the system 
software (e.g. compiler and library versions) and on the run-time environment. 
The latter may -- and usually do -- change during the months of production. Therefore,
the times quoted here can only serve as an indication of the approximate
cost of the simulations and have to be taken with care.

\setlength{\tabcolsep}{1pt}
\begin{table}
\begin{center}
\small
\begin{tabular}{@{\extracolsep{3mm}}ccccccccccc}
\toprule
id 	& machine & $V_{\rm local}$ & $N_{\rm cores}$ & $N_{\rm ppc}$ & min/MDU\\
\midrule
H101r000 & SuperMUC & $8\times4\times8^2$   &1536 &1 &  9 \\ 
H102r002 & SuperMUC &  $8\times4\times8^2$  &1536 &1 &  8 \\ 
H105r002 & SuperMUC &  $8\times4\times8^2$  &1536 &1 & 10 \\ 
C101r013 & SuperMUC &  $8^4$                &2592 &1 & 27 \\ 
\midrule
H200r000 & SuperMUC &  $8\times4\times8^2$  &1536 &1 &  8 \\
N200r000 & SuperMUC &  $8\times12\times6^2$ &4096 &1 & 12 \\
D200r000 & JUQUEEN  &  $8\times4^2\times8$  &8192 &4 & 59 \\
\midrule
N300r002 & SuperMUC &  $8^2\times6\times12$ &3072 &1 & 13 \\
J303r003 & FERMI    &  $12\times4^3$        &16384 &4 & 33 \\ 
\bottomrule
\end{tabular}
\caption{\label{tab:cost}Production setup of selected runs. The last 
column shows the wall-clock time in minutes per molecular dynamics unit 
on the specific machines used in this project. 
The other columns give the local lattice size per 
MPI process ($V_{\rm local}$), the number of cores used ($N_{\rm cores}$), and
the number of MPI processes run on each core ($N_{\rm {ppc}}$). Since execution
times also depend on the actual system software and on the run-time environment, 
the last column can provide only a rough indication of the cost of the simulations.}
\end{center}
\end{table}

\section{Autocorrelations\label{sec:ac}}
Markov Chain Monte Carlo algorithms, like the Hybrid Monte Carlo used here,
produce field configurations which exhibit
autocorrelations characterized for an observable $A$ by the autocorrelation
function
\begin{equation}
\Gamma_A(t) = \langle A_t\,A_0 \rangle - \langle A \rangle^2 \,,
\end{equation}
where $t$ is the Monte Carlo time. 
The integral over the
normalized autocorrelation function $\rho(t)$ enters the error analysis. This is the
integrated autocorrelation time
\begin{equation}
\tauint(A)=\frac{1}{2} + \sum_{t=1}^\infty \rho_A(t) \equiv \frac{1}{2} + \sum_{t=1}^\infty \frac{\Gamma_A(t)}{\Gamma_A(0)} \,.
\end{equation}
To estimate  $\tauint(A)$ with a finite variance, it is necessary to cut
the summation at a window $W$~\cite{Madras:1988ei,Wolff:2003sm}.
In order to  choose the window for our final error estimates and to account for the thereby 
neglected tail, we employ the method described in \Ref{Schaefer:2010hu}.
Its essential input is an  estimate for the exponential autocorrelation time, which
we discuss in the following.

\subsection{Scaling of the autocorrelations}

As we approach the continuum limit, the autocorrelation times are expected to
grow due to critical slowing down. The open boundary conditions used in our
setup should prevent catastrophic scaling due to the freezing of the
topological charge.   Since we have chosen the trajectory length constant in all
our runs, we expect Langevin scaling $\tauint \propto a^{-2}$.

\begin{figure}
\begin{center}
\includegraphics[width=0.6\textwidth]{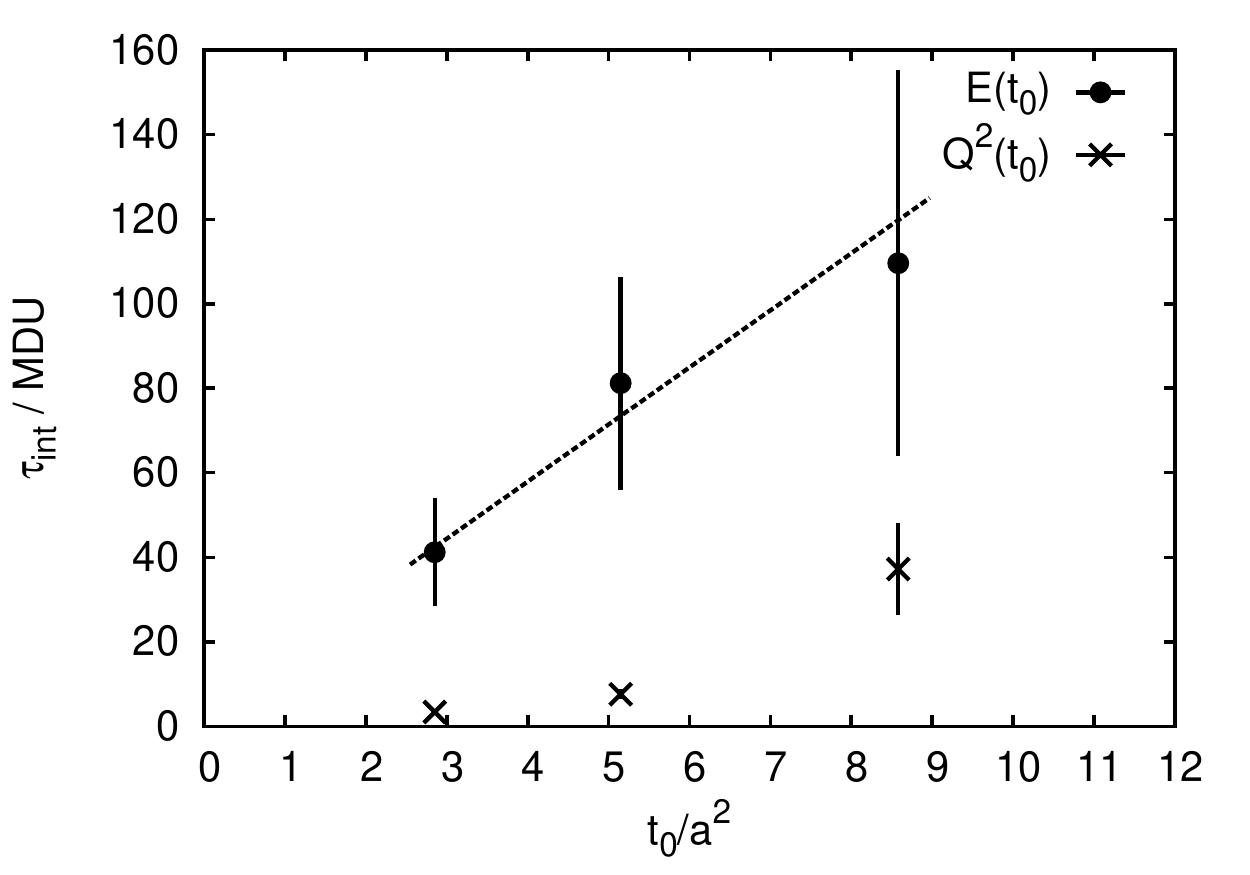}
\end{center}
\caption{Scaling of the integrated autocorrelation time of $Q^2(t_0)$ and $E(t_0)$.
For the energy, we observe very good scaling, whereas for the charge, significant violations are observed.
At coarser lattices the topological charge decorrelates significantly faster than predicted by the scaling
hypothesis, very similar to the pure gauge case \cite{Luscher:2011kk}. \label{fig:ac}}
\end{figure}

In \Fig{fig:ac} we show autocorrelation times of notoriously slow observables:
the global topological charge and the action density averaged over the plateau region,
both constructed from links smoothed by the Wilson flow integrated to flow time $t_0$. They are both
defined in \Eq{eq:EQ}. We find a
situation similar to that encountered in pure gauge theory~\cite{Luscher:2011kk}. While the 
energy density shows very good scaling, the topological charge decorrelates 
faster on coarser lattices. 

The fast growth of the integrated autocorrelation time of the charge does not
mean that the $1/a^2$ scaling is not valid. In pure gauge theory, the behavior
could very well be fitted with $\tauint \propto a^{-2}( c+d a^2)$. In this picture,
there are significant cutoff effects to the scaling, but no catastrophic behavior
in the $a\to0$ limit. This is expected when simulating with open boundary conditions.

\subsection{Cost of the simulation}

Note, in two-flavor QCD with periodic boundary conditions at a lattice spacing 
of roughly $a=0.05\,\fm$ \cite{Bruno:2014ova} the topological charge does not decorrelate slower
than the smoothed energy. Rather, it shows similar autocorrelations for quark masses around $400\,\mev$.
This means that we are not yet in the position to
fully profit from the effect of the open boundary conditions, however, going to
finer lattice spacings the freezing observed in two-flavor QCD at $a\approx
0.03\,\fm$~\cite{Schaefer:2009xx} will be avoided.

In the sense of fast decorrelations and minimal requirements on the number of
units of molecular dynamics time (MDU), the presented simulations are not
cheap, nevertheless. The exponential autocorrelation time of $\tauexp\approx
14(3)\, t_0/a^2$ is consistent with what is found in \fig{fig:ac}. For biases
to be small and a simulation to be reliable we need a total Monte--Carlo
history of  at least $\rmO(50)\times \tauexp$. For $\beta=3.4$ this
translates to $2000\,\mathrm{MDU}$, whereas for $\beta=3.55$ and $\beta=3.7$ a
statistics of $3600\,\mathrm{MDU}$ and  $6000\,\mathrm{MDU}$, respectively,  is
necessary. For most of our ensembles listed in \Tab{tab:ens}, we exceed these
numbers, but for some, which are still in production, they are not yet reached.
Those quoted results therefore have to be taken with care in these cases.

\section{Reweighting factors\label{sec:rw}} 

The simulations are not done with the exact QCD action as given by
\Eqs{eq:gauge} and (\ref{eq:ferm}), but differ due to the twisted-mass
reweighting \Eq{eq:effmf} and the inaccurate rational function in the RHMC
\Eq{eq:effrhmc}. The observables are reweighted to the target theory, for which the
reweighting factor $W=W_0W_1$  needs to be computed. The factors $W_0$ and $W_1$, as defined in \Eqs{eq:w0} and
(\ref{eq:rhmc}) respectively, contain ratios of determinants which are
estimated stochastically as described below.

Expectation values
$\langle A \rangle$ of primary observables $A$ can then be computed from
expectation values in the theory with the modified action $\langle \cdots
\rangle_W$, according to
\begin{equation} 
\langle A \rangle = \frac{\langle A W \rangle_W}{\langle W \rangle_W}\,. 
\end{equation}

\subsection{Twisted-mass reweighting factor\label{s:tmrw}}

The twisted-mass reweighting plays an important role in our setup. From a
conceptual point of view, it removes barriers of infinite action created by
zero eigenvalues of the Wilson Dirac operator. Together with Hasenbusch
factorization, it also reduces the fluctuations of the forces which makes the
simulations cheaper and more reliable in practice~\cite{Schaefer:2012tq}.

This situation is especially favoured for a large value of $\mu_0$ in
\Eq{eq:w0}. It might, however, also lead to significant fluctuations in the
reweighting factor and as a consequence to a larger statistical error of
certain observables.

As a consequence, what constitutes the optimal choice of the parameter $\mu_0$
will in general depend on the observable. As can be seen in \Fig{fig:rw}, $W_0$
is close to a constant for most configurations; only on some configurations the
value will be much smaller. For observables with little or no correlation to
the reweighting factor, like the gluonic ones we consider below, this
effectively amounts to a reduction in statistics~\cite{Bruno14}. This reduction
is negligible for our ensembles since $\langle\mathrm{var}(W)\rangle \ll
\langle W \rangle^2$ in all cases.

For observables with a strong correlation  with $W_0$, the    situation is more 
delicate. Even after reweighting, this can lead to large fluctuations in the measurements
and significantly increased statistical error. In particular in the case of anticorrelation,
the situation is  more problematic due to the stochastic estimation of $W$ and, possibly,
the observable. The cancellation between, e.g. a large value of the observable and a small 
value of $W$ might require a rather precise determination of the two.

\subsection{Reweighting and the pseudoscalar correlation function}

The pseudoscalar correlation function  is an observable showing a strong
anticorrelation between its value and the reweighting factor. This can be 
easily understood by noting that at small quark masses both receive significant
contributions from the smallest (in magnitude) eigenmodes of the Hermitian
Dirac operator. It is precisely this region where the reweighting term has the 
largest effect.

\begin{figure}
\begin{center}
\includegraphics[width=\textwidth]{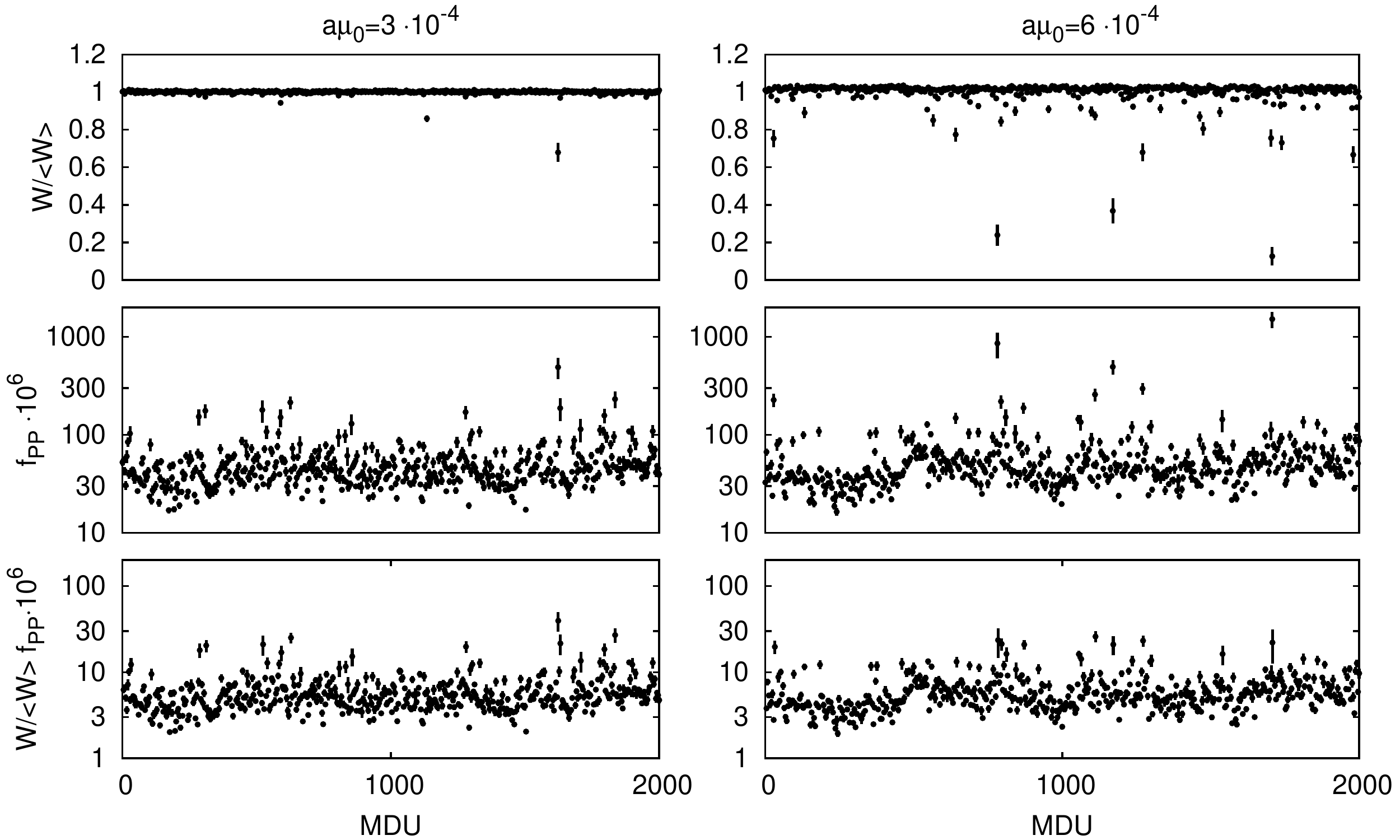}
\end{center}
\caption{Time history of the reweighting factor (top),
the pseudoscalar correlator $f_\mathrm{PP}(x_0)$ (center) and
the product $Wf_\mathrm{PP}(x_0)$ (bottom) on two C101 ensembles 
with different values of the reweighting parameter, $a\mu_0=0.0003$ and  $a\mu_0=0.0006$, respectively,
evaluated on time-slice $x_0=(T+a)/2$. The error bars indicate the uncertainty due
to the stochastic estimation of these quantities.\label{fig:rw}}
\end{figure}

To illustrate the cancellation between the fluctuations in $W$ and
$f_\mathrm{PP}(x_0)$, \Fig{fig:rw} displays the time series of the two (top
and central panel)  at $x_0=(T+a)/2$ together with the product
$Wf_\mathrm{PP}(x_0)$; see \Eq{eq:fpp} for its  definition. Data for C101 and
two values of $\mu_0$ is shown. As we can see, the larger $\mu_0$ leads to
larger fluctuations in $W$ and $f_\mathrm{PP}(x_0)$, as expected. In the
product, however, they cancel and the average value $\langle W
f_\mathrm{PP}(x_0) \rangle / \langle W \rangle $ is then consistent within the
statistical errors between the two ensembles.

\subsection{Computation of  \texorpdfstring{$W_0$}{}}
Since the determinant ratios needed for the computation of $W_0(\mu_0)$
cannot be computed directly, a stochastic estimator is taken instead.
This can either be done by directly  estimating the determinant ratio 
in \Eq{eq:w0} or by first splitting it up and then using stochastic estimates 
for the individual factors~\cite{Hasenfratz:2008fg}, 
a strategy already successful in Hasenbusch's mass factorization.

Among the many possibilities, we here restrict ourselves to splitting the 
interval  between $\mu=0$ and $\mu=\mu_0$ into $N_\mathrm{sp}$ smaller steps
$\mu_0=\tilde \mu_0>\tilde \mu_1>\dots>\tilde \mu_{N_\mathrm{sp}}=0$
\begin{align}
W_0(\mu_0)=\prod_{i=1}^{N_\mathrm{sp}}\det R(\tilde \mu_{i-1},\tilde \mu_{i})\,; &&
R(\mu_1,\mu_2)= \frac{(\hat Q^2 + \mu_2^2)^2}{(\hat Q^2+ \mu_1^2)^2} \frac{(\hat Q^2 + 2  \mu_1^2)}{(\hat Q^2 + 2 \mu_2^2)}\,.
\end{align}
Now each of the factors is evaluated stochastically with $N_\mathrm{r}$ 
complex-valued Gaussian random fields $\eta$ of unit variance
\begin{equation}
\tilde R(\mu_1,\mu_2,N_\mathrm{r})= 
\frac{1}{N_\mathrm{r}} \sum_{i=1}^{N_\mathrm{r}} 
\mathrm{exp}\big \{-\left(\eta_i,(R^{-1}(\mu_1,\mu_2)-1) \eta_i\right)\big\}\,,
\end{equation}
such that up to an irrelevant constant factor the determinant is retrieved by averaging
over the noise fields
\begin{equation}
\det R(\mu_1,\mu_2) \propto \langle \tilde R(\mu_1,\mu_2,N_\mathrm{r}) \rangle_\eta.
\end{equation}

Following the initial proposal of \Ref{Luscher:2012av}, it is  sufficient to
use a single step $N_\mathrm{sp}=1$  with a suitably chosen value of
$N_\mathrm{r}$. Its value along with the other parameters of the reweighting
can be found in \Tab{tab:rew}. This is the method implemented in openQCD-1.2. 

Once the fluctuations in the reweighting factor increase, it is advisable to
use intermediate $\tilde \mu$, a possibility given in openQCD-1.4. This is
because the distribution of the results for the reweighting factors become
long-tailed once exceptionally small eigenvalues of the $\hat Q^2$ are
encountered. In this situation it is very difficult to argue about the
uncertainty of $W_0$~\cite{Hasenfratz:2002ym}. By splitting the estimate into
smaller intervals in $\tilde \mu$, the distribution of each of the factors
becomes significantly more regular.

For ensemble H105r002 we find precisely such a situation. While with a single
step in $\tilde \mu$ the smallest reweighting factors show a distribution which
is far from Gaussian,  using ten intermediate $\tilde \mu$ the individual factors
can be computed reliably to $\mathrm{O}(15\%)$ accuracy using 15 sources each.

\begin{table}
\small
\begin{center}
\begin{tabular}{@{\extracolsep{4mm}}cllccc}
\toprule
id  & $N_\mathrm{r}$& $\frac{\mathrm{var}(W_0)}{\langle W_0 \rangle^2} \cdot 10^3$ & $\frac{\mathrm{var}(W_1)}{\langle W_1 \rangle^2} \cdot 10^5$   \\
\midrule
H101     & 12          & 0.00047(9) & 5.1(2)       &   \\ 
H102     & 12          & 0.036(4)   & 1.88(5)      &   \\ 
H105     & 36          & 3.2(4)     & 7.3(2)       &   \\ 
H105r005 & 24          & 0.0032(9)  & 3.7(2)       &   \\ 
C101     & 24          & 1.8(1.1)   & 1.6(2)       &   \\ 
C101r014 & 24          & 5.1(2.1)   & 1.63(10)     &   \\ 
\midrule               
H200     & 24          & 0.00018(5) & 4.7(2)       &   \\ 
N200     & 24          & 0.4(2)     & 2.23(7)      &   \\ 
D200     & 48          & 0.15(5)    & 4.9(3)       &   \\ 
\midrule               
N300     & 24          & 0.00018(2) & 3.0(1)      &   \\ 
\sl J303 & \sl 24  & \sl 3.7(3.2) & \sl 1.3(2) &   \\
\bottomrule
\end{tabular}
\end{center}
\caption{Parameters of the reweighting. We give the number of sources $N_\mathrm{r}$ used to estimate the twisted-mass
reweighting factor $W_0$ --- for the RHMC reweighting factor $W_1$ we always use one source --- and the resulting variances of $W_0$ and $W_1$. $N_\mathrm{sp}=1$ in all cases. H105 refers
to runs r001 and r002, whereas C101 to runs r013 and r015.   J303 have not reached sufficient 
statistics for a reliable result.\label{tab:rew}}
\end{table}

\subsection{RHMC reweighting factor}

Since the rational approximation has been chosen to a good accuracy, the
fluctuations in the reweighting factor are small and it turns out to be
sufficient to estimate it with one stochastic source per configuration. 
The associated variances are  given in \Tab{tab:rew}. They are seen to receive a considerable
contribution from the stochastic estimation of $W_1$. 

In order to study the effect of more sources,
we observe using five instead of one stochastic estimate
reduces the variance of $W_1$ by more than a factor 4, on ensemble H105r005. The same is true for the H200 ensembles.
Still, even with one source per gauge configuration 
the noise introduced by $W_1$ is negligible for all observables we investigated.

Note that in some early runs we underestimated the upper bound of the interval in
which the rational function is accurate. Since the accuracy does not
deteriorate quickly outside the interval, the fluctuations of the reweighting
factors nevertheless are sufficiently small.

\section{Observables\label{sec:meas}}
\subsection{Wilson flow}

The Wilson flow can be a very useful
tool in lattice QCD from which quantities with a finite continuum limit can
be constructed~\cite{Narayanan:2006rf,Luscher:2010iy,Luscher:2011bx}. The gauge fields $U(x,\mu)$  are
subjected to the smoothing flow equation
\begin{equation}
\partial_t V_t(x,\mu) = -g_0^2 \{ \partial_{x,\mu} \SW (V_t) \} V_t (x,\mu) 
\,, \quad V_t(x,\mu)\big|_{t=0} = U(x,\mu)\,,
\end{equation}
with $\SW$ being the Wilson action. With clover-type discretization
of the field strength tensor $\hat G_{\mu\nu}(x,t)$ constructed from the smooth fields $V_t$,
the time slice energy  $E(x_0,t)$ and the global topological charge $\Qtop(t)$ can be constructed
\begin{equation}
\begin{split}
E(x_0,t)&=-\frac{a^3}{2L^3}\sum_{\vec x}  \mathrm{tr} \{ \hat G_{\mu\nu}(x,t) \,  \hat G_{\mu\nu}(x,t) \} \,,\\
\Qtop(t)&=-\frac{a^4}{32\pi^2}\sum_x\, \epsilon_{\mu\nu\alpha\beta}\, \mathrm{tr} \{ \hat G_{\mu\nu}(x,t) \,  \hat G_{\alpha\beta}(x,t) \} \,.
\end{split}
\label{eq:EQ}
\end{equation}

With the vacuum expectation value of the energy $\langle E(t) \rangle$, the scale
parameter $t_0$ is then defined by
\begin{equation}
t^2 E(t) \big |_{t=t_0} = 0.3 \,.
\label{eq:t0}
\end{equation}
Throughout this paper we quote the observables of \Eq{eq:EQ} at flow time $t_0$.

\subsection{Effects of the boundary}
Due to the open boundary conditions in the temporal direction, time translational
invariance is lost. Sufficiently far away from the boundaries, local observables
are expected to assume their vacuum expectation values up to exponentially small
corrections with a decay rate equal to the lightest excitation with vacuum quantum
numbers.

\begin{figure}
\begin{center}
\includegraphics[width=0.65\textwidth]{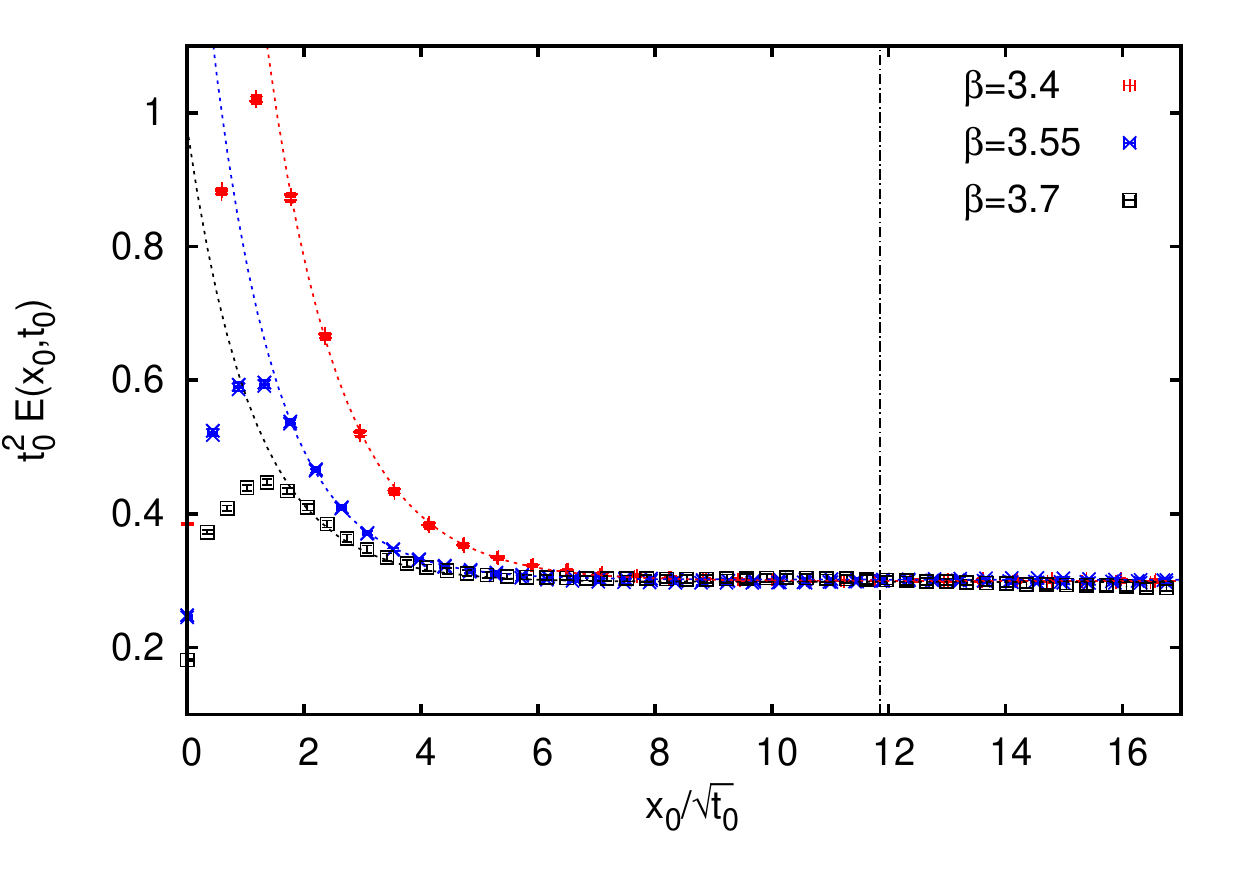}
\end{center}
\caption{In the vicinity of the boundary, significant cutoff effects in $E(x_0,t_0)$ are observed. They 
are noticable in the expected region: $x_0=4a$ is roughly at $x_0/\sqrt{t_0}=2.3$, $1.7$ and $1.4$ for $\beta=3.4$, $3.55$ and
$3.7$, respectively. At the 
same time, the dependence on the quark masses is negligible: for each lattice spacing we plot the data for all
available quark masses. The dotted lines represent fits to \Eq{eq:bndEt}. They are 
used to set the lower bound of the plateau fit indicated by the vertical dashed line.\label{fig:bnd}}
\end{figure}

On top of this continuum boundary effect, large discretization errors are observed
close to the boundary. As an example we show in \Fig{fig:bnd} the behavior of the
smoothed energy $E(t,x_0)$, defined in \Eq{eq:EQ}. 
A further example for the pseudoscalar correlation function with the sink
approaching  the boundary can be found in \Ref{Bruno14}.

In the case of the energy, it should be noted that it is at
this point difficult to disentangle the discretization effects in the
underlying gauge field from the ones introduced by the Wilson flow and the
observable used to define the field strength tensor, but recent work by Ramos and Sint
might clarify this issue~\cite{Alberto14}. Also the Dirac operator
used in the measurement of the effective mass is only tree-level improved at
the boundary.  Still, the effects of the finite lattice spacing are very
prominent at our coarser lattices, but become much less notable as the continuum limit is
approached. 

As can be seen in \Fig{fig:bnd}, no sizable dependence on the quark mass  is
observed in $t_0^2 E(x_0,t_0)$. This is trivial for the bulk, since its value
is equal to 0.3 by definition. But also the (cutoff) effects close to the
boundaries show no quark mass dependence. Whether this is a generic feature of
the sea quark contribution being small or it is due to our particular choice of
chiral trajectory \Eq{eq:chitra} cannot be judged from the data presented here.

In the context of the present paper we will not discuss these effects in
detail, but perform the measurements in a region where they can be neglected.
The determination of the plateau region is not always clear due to an effect
already observed in \Ref{Luscher:2012av}: In precise observables, like the
examples above, long-range waves are visible. They are a consequence of the limited
statistics and do not exceed what is expected if the statistical analysis is
done properly, however, they do make the plateau determination more difficult. 

For meson correlation functions, these waves have been discussed
previously \cite{Aoki:1995bb}, see \Fig{fig:meff} for an example from our
simulations. In other simulations they are typically not visible, because time
translational invariance is used on the level of the correlation function by
using different source  positions in time and averaging them before computing
effective masses. Again, this is not a principal problem. However, we need to 
 ensure sufficient statistics and that errors are under control and  
require a procedure to deal with these waves.

\subsection{Measurement of \texorpdfstring{$t_0$}{}\label{sec:t0}}
For the determination of $t_0$, we need to determine the plateau in $E(x_0,t)$ for $t=t_0$. 
Since we are looking at a smoothing radius of $\sqrt{8 t_0}$, the effects of the boundary
visible in  \Fig{fig:bnd} are at the expected length scale. Discretization effects are large though,
and it is therefore difficult to argue about the expected functional form.
In this situation, we use a two-stage procedure: First we fit
\begin{equation}
E(x_0,t)=E(t)+c_0 e^{-m\frac{T}{2}} \cosh\{-m (x_0-\frac{T}{2}) \}
\label{eq:bndEt}
\end{equation}
in the range where this ansatz describes the data. This is only used to determine the fit range
by the condition that in the whole range the statistical uncertainty $\delta E(x_0)$ is dominating over the
systematic effect from a non-vanishing $c_0$, i.e., we require for the fit interval $[x_{0,\mathrm{min}},T-x_{0,\mathrm{min}}]$
\begin{equation}
\frac{1}{4}\,\delta E(x_{0,\mathrm{min}},t) > c_0 e^{-m\frac{T}{2}} \cosh\{-m (x_{0,\mathrm{min}}-\frac{T}{2})  \} \,.
\end{equation}
At the current accuracy of the data, the result of this investigation is that a
single $x_{0,{\rm min}}$ is sufficient for each value of $\beta$, as might be
expected from \Fig{fig:bnd}. The effect of the quark mass is negligible. In
particular we have
\begin{align}
  x_{0,\mathrm{min}}(\beta=3.4)/a&=20 \,; &
  x_{0,\mathrm{min}}(\beta=3.55)/a&=21\,; &
  x_{0,\mathrm{min}}(\beta=3.7)/a&=24  \,,
\end{align}
and the final value of $E(t)$ in the vicinity of $t=t_0$ is determined by
averaging $E(x_0,t)$ in the corresponding interval.
The value of $t_0/a^2$ is then determined by \Eq{eq:t0}. The results are listed in \Tab{tab:mpst0}.

\begin{figure}
\begin{center}
\includegraphics[width=0.65\textwidth]{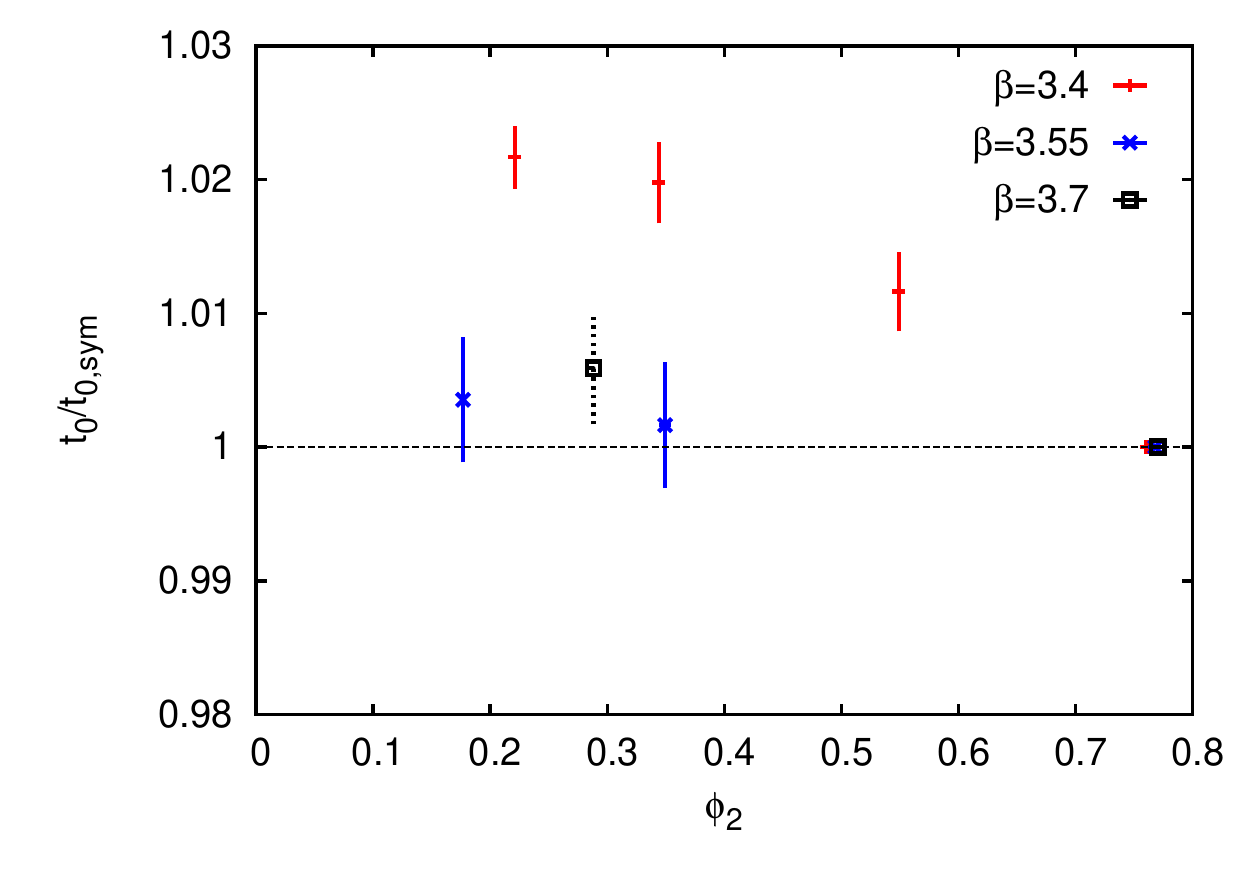}
\end{center}
\caption{Dependence of $t_0/t_{0,\mathrm{sym}}$ on $\phi_2$, where $t_{0,\mathrm{sym}}$ is the value at
$m_\mathrm{ud}=m_\mathrm{s}$ along our trajectory. The dashed errorbars indicate the low statistics in the J303 ensemble. \label{fig:t0phi2}}
\end{figure}

In \Fig{fig:t0phi2} the quark mass dependence of $t_0$ is given for the three available
lattice spacings. Recall, the values are given in terms of the symmetric point
which defines the chiral trajectory at $\mud=\ms$. From a  Taylor expansion
around this point~\cite{Bietenholz:2010jr} as well as ChPT~\cite{Bar:2013ora}
one expects a constant behavior to the respective leading order. This is
confirmed for the finer lattices to our level of accuracy. Only at the coarsest
lattice spacing, cutoff effects seem to cause some deviation, albeit on a
rather small scale.

\subsection{Pseudoscalar masses}
The masses of the pseudoscalar particles are computed from the 
pseudoscalar correlation function projected to zero momentum. 
With quark fields of flavor $r$ and $s$, and the pseudoscalar 
density $P^{rs}=\bar \psi^r \gamma_5 \psi^s$, it is given by
\begin{equation}
f_\mathrm{PP}(x_0,y_0)=-\frac{a^6}{L^3}  \sum_{\vec x,\vec y} \langle P^{rs}(x) P^{sr}(y) \rangle \ .
\label{eq:fpp}
\end{equation}
Due to the open boundary conditions in time, the translational
invariance in the temporal direction is broken. However, we find 
that there is little to gain from using source fields at different
time slices~\cite{Bruno14}. The $U(1)$ stochastic source fields
are therefore put only at $y_0=a$ and $y_0=T-a$~\cite{Sommer:1994gg}. 
In the following we analyze 
\begin{equation}
f_\mathrm{PP}(x_0)\equiv \frac{1}{2}\big \{f_\mathrm{PP}(x_0+a,a)+f_\mathrm{PP}(T-a-x_0,T-a)\big\}\,.
\end{equation}

In the continuum limit and for large volume and  sink positions far away from the source and
boundary, $x_0\gg 0$ and $x_0\ll T$, the two-point function is expected to fall off as~\cite{Luscher:2012av}
\begin{equation}
f_\mathrm{PP}(x_0) = A\, \sinh\big (m_\mathrm{PS} (\tilde T-x_0)\big ) \,.
\label{eq:fpp1}
\end{equation}
In line with \Ref{Luscher:2012av}, $\tilde T$ is a free parameter. We follow a similar
strategy as in \Sect{sec:t0} to make sure that in our final fit the excited state contribution
is negligible.

\begin{figure}
\begin{center}
\includegraphics[width=.49\textwidth,clip]{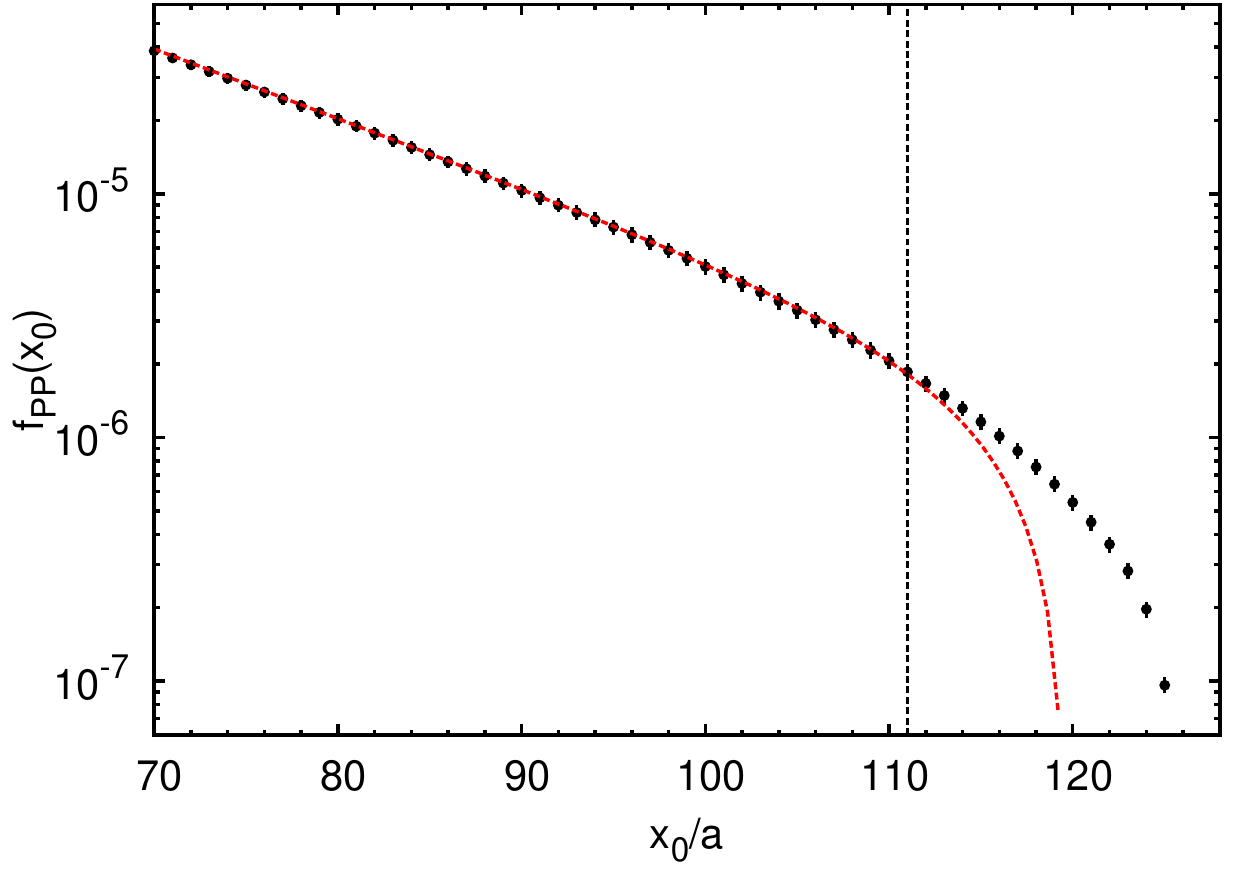}
\includegraphics[width=.49\textwidth,clip]{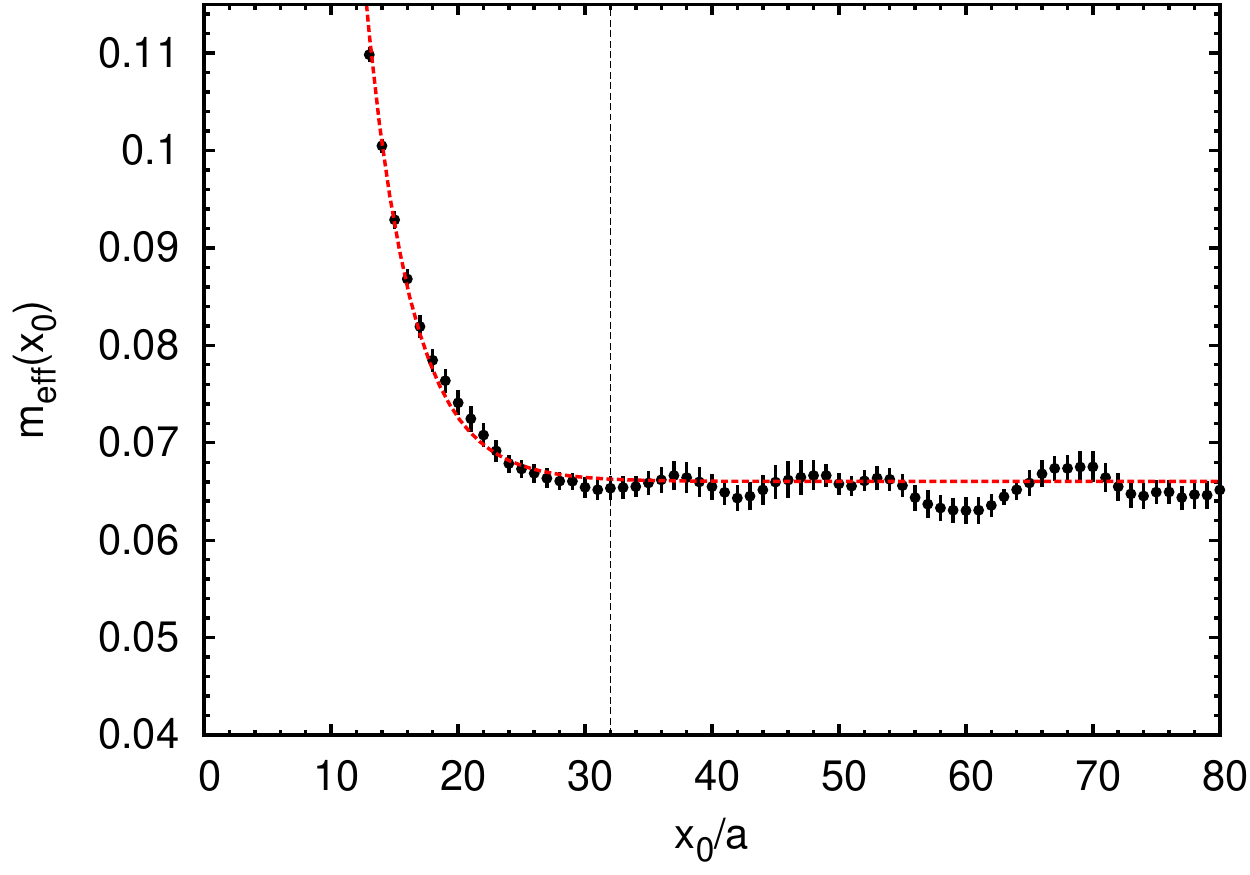}
\end{center}
\caption{Left: Pion propagator on the D200 lattice with the line representing a fit
to \Eq{eq:fpp1} and the vertical line giving the upper bound of the fit interval. Right:
Effective mass of the same correlation function, with the line representing a fit
including one excited state.\label{fig:meff}}
\end{figure}

We show an example of an effective mass plot in \Fig{fig:meff}, where we can see that this fit works
very well in a wide range of $x_0$.
The  results for the masses are listed in \Tab{tab:mpst0}.

\begin{table}
\begin{center}
\small
\begin{tabular}{@{\extracolsep{3mm}}cccccc}
\toprule
id & $am_\pi$ & $am_K$ & $t_0/a^2$ & $\phi_2$ & $\phi_4$\\
\midrule
H101 & 0.18273(70) & 0.18273(70) & 2.8468(61) & 0.7605(44) & 1.1407(77) \\ 
H102 & 0.15437(70) & 0.19164(57) & 2.8799(73) & 0.5490(44) & 1.1206(69) \\ 
H105 & 0.12170(96) & 0.20126(63) & 2.9031(73) & 0.3440(49) & 1.1127(80) \\ 
C101 & 0.09751(93) & 0.20639(40) & 2.9085(51) & 0.2212(38) & 1.1017(54) \\ 
\midrule 
H200 & 0.13653(52) & 0.13653(52) & 5.150(23)  & 0.7680(60) & 1.1520(88) \\  
N200 & 0.09202(61) & 0.15059(57) & 5.1584(78) & 0.3494(48) & 1.1105(88) \\ 
D200 & 0.06542(44) & 0.15640(25) & 5.1681(68) & 0.1769(26) & 1.0998(39) \\ 
\midrule
N300 & 0.10593(32) & 0.10593(32) & 8.580(27) & 0.7702(53) & 1.1553(79) \\
\sl J303 & \sl 0.0648(3) & \sl 0.1198(3) & \sl 8.63(3)  & \sl 0.288(3) & \sl 1.136(6) \\

\bottomrule
\end{tabular}
\caption{Measured values for the pseudoscalar masses, the scale $t_0/a^2$ and the two scaling
variables $\phi_2$ and $\phi_4$. The results in C101 are based on runs r013, r014 and r015. 
J303 has a statistics of roughly $20\tau_\mathrm{exp}$. The values on this ensemble are therefore
not reliable. \label{tab:mpst0}}
\end{center}
\end{table}

Even though we do not give results on decay constants, let us remark that also
in this case the sources can be put in the vicinity of the boundaries. Methods
similar to the ones already developed in the Schr\"odinger Functional
\cite{Guagnelli:1999zf} can be used to cancel the matrix element of the source
operator such that only the sink has to be sufficiently far away from the
boundaries. Various possibilities for open boundary conditions on  
the ensembles presented  are discussed in \Ref{Bruno14}.

\subsection{Comparison to simulations with periodic boundary conditions}
One possible concern regarding the open boundary conditions is that the region close
to the boundaries is large and as a consequence one loses a sizeable fraction of the 
statistics. Which fraction of the lattice needs to be discarded depends on the observable
and the statistical accuracy of the data, with boundary effects expected to decay close to 
the chiral limit as $\exp(-2m_\pi x_0)$. 
However, it should be noted that also the systematic finite volume effects are parametrically
similar with contributions proportional to  $\exp(-m_\pi L)$: A high accuracy requires
large lattices in both, the temporal and spatial directions, which is also true for
simulations with periodic boundary conditions in time. 

In any case, since we observe large cutoff 
effects close to the boundary, arguments based on continuum physics are problematic at
current lattice spacings.
This is already visible in the $x_{0,\mathrm{min}}/a$ chosen in the measurement
of $E(t_0)$, which has only a minor dependence on the lattice spacing. On our
smallest lattices at $\beta=3.4$ with $N_\mathrm{t}=96$, the
$x_{0,\mathrm{min}}/a=20$ leads to a plateau average of almost $60\%$ of the
total time extent. On all other ensembles we have an even larger fraction over which
we can take the plateau average.

In our measurements of the pseudoscalar masses we typically start the plateau at 
$x_{0,\mathrm{min}}\approx T/4$, from where on the effects from the excited states
can be neglected. As noted before, moving the source away from the boundary has little effect, since
the plateau is seen to start at the same position. The minimal distance $x_{0,\mathrm{max}}$  of the sink from the
boundary is typically around $T/6$, such that we have in total 
a plateau stretching between 50\% and 65\% of the lattice. Even if the other half 
of the lattice was completely decorrelated, this would at most correspond to a factor of two
in statistics.

\section{\label{sec:beta33}Scaling violations}
In the bulk, our action is fully  $\rmO(a)$ improved, only for 
the boundary terms we use the tree-level values. This guarantees leading
scaling violations close to the continuum limit to be of order $a^2$,
but at finite lattice spacings higher order terms will always be present as well.
How large their contribution is and whether one can safely neglect them
given the statistical uncertainties of the simulation is not a priori clear.
In any case, once the higher order terms become important, they limit the value
of coarser lattices in the continuum extrapolation.

\subsection{Cutoff effects in \texorpdfstring{$t_0$}{}\label{sec:t0cut}}

A particularly precise way to study discretization effects is to look at
observables which agree in the continuum limit but differ at finite lattice
spacing. To this end, we take two slightly different definitions of $t_0$: both
are given by the implict relation \Eq{eq:t0}, where in one case we use the
conventional ``clover'' discretization of the field strength tensor for the 
the energy density  $E$, \Eq{eq:EQ}, in the other case the plaquette definition
is used as given in \Ref{Luscher:2010iy}. 

Since the continuum value of $t_0$ has to be the same, the ratio of
$t_0^\mathrm{clov}$  and $t_0^\mathrm{plaq}$ has to be one up to cutoff effects.
As they are evaluated on the same gauge field configurations, the two values of $t_0$
are highly correlated such that their ratio can be evaluated to exceedingly high accuracy.

\begin{figure}
\begin{center}
\includegraphics[width=0.6\textwidth]{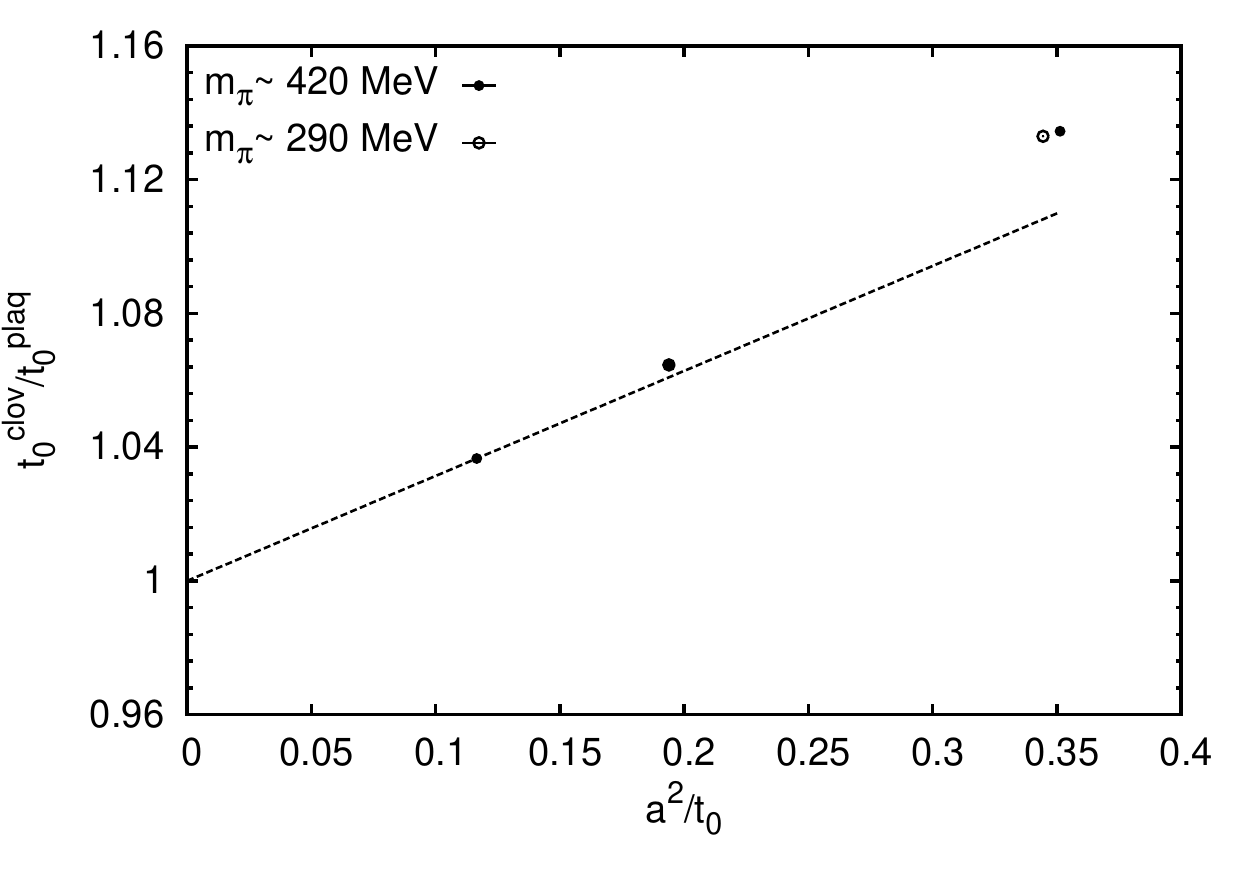}
\end{center}

\caption{\label{fig:t0a2} Ratio of the values of $t_0$ from two different
discretizations (plaquette and clover) of the field strength tensor used in
$E$, which is constructed from the same smoothed gauge fields at flow time
$t_0$ as given in \Tab{tab:mpst0}. From left to right we show the values at
$\beta=3.7$, 3.55 and 3.4, with the straight line given by the continuum value
and the point at $\beta=3.7$. The point at $\beta=3.3$ is not shown, it lies at
$a^2/t_0=0.5$ with the ratio on the y-axis at about 1.3.}

\end{figure}

As we can see in \Fig{fig:t0a2}, the ratios at $\beta=3.7$ and $\beta=3.55$
agree with the  $a^2$ scaling hypothesis up to very high accuracy, with a total
deviation of the ratio from its continuum value of $4\%$ and $6\%$,
respectively.  With the assumption that higher order effects are negligible at $\beta=3.7$,
one concludes that at $\beta=3.55$ higher orders contribute $0.4\%$ to this observable,
while at $\beta=3.4$ an
additional $\mathrm{O}(2\%)$ effect can be attributed to higher orders on top
of the $11\%$ which come from the leading order scaling violation. 

While this additional $\mathrm{O}(2\%)$ effect originating from the higher order terms is not
of concern for most observables in current lattice computations, it might 
impact studies of certain high accuracy observables.

In any case, the large discrepancies between the two definitions of $t_0$ are a
problematic finding in view of the fact that our tuning strategy is entirely
based on this quantity to set the scale. We therefore have to expect that for
other scale setting strategies, the matching of the chiral trajectories will
differ on the order of the ratio observed at the level of $10\%$ at the
coarsest lattice spacing.

\subsection{Coarser lattices}

In order to investigate the value of ensembles at coarser lattice spacings, we
have also generated some $\beta=3.3$ lattices at the $\mud=\ms$ matching point.
After some tuning,  $\kud=\ks= 0.136423 $ is  found to match the $\phi_4=1.15$
point to reasonable accuracy. However, the observation of large cutoff effects
on $96\times 24^3$ lattices, indicating this  point no longer being in the
assumed $a^2$ scaling region, has led us to abandon this coupling for now.

This decision is based on \Fig{fig:t0a2}, where for this parameter point we find
$a^2/t_0\approx 0.5$ and a ratio of  $t_0^\mathrm{clov}/t_0^\mathrm{plaq}\approx
1.3$, which is $16\%$ above the leading order violations of 1.15. We are
therefore clearly no longer in the scaling regime and since we aim with  most
observables at accuracies much below the $10\%$ level, the points at this
lattice spacing of roughly $0.1\,\fm$, do not meet our precision goals.

Furthermore, the autocorrelations observed in particular in the thin link
plaquette were very significant and also large fluctuations in the lower
spectrum of the Dirac operator have been observed.  This makes these lattices
difficult to simulate and poses another reason to refrain from considering this
value of $\beta$ at this time.

\section{Conclusions}
The generation of the gauge field configurations described here lays the ground
for many future lattice QCD calculations. It is the first time that open
boundary conditions in time and twisted-mass reweighting have been
extensively used in such large scale calculations.

The two methods have been shown to work well. Keeping in mind simulations which
are on our roadmap for the future, the experience gained with the use of open
boundaries will prove very valuable as the lattice spacing is decreased, while
not being strictly necessary at the lattice spacings under investigation.

We could show, that no particular obstacle is posed by the boundaries
themselves, however, significant discretization effects are observed in their
vicinity. Depending on the observable and its correlation with the reweighting
factor, we observe the twisted-mass reweighting is under control. To study this
in the future, we have generated ensembles with different values of the
reweighting parameter $\mu_0$.

It is noteworthy that similar data sets previously needed to be accumulated
over many years. However, due to advances in hardware and in algorithms, we
could demonstrate the progress that has been made, by generating the current,
new data set within a year and a half after the parameters of the action had
been determined.

As of now, the covered parameter space is limited: We only have data on one
chiral trajectory, a limited range of quark masses and lattice spacings and
typically only one volume. In order to better control the associated systematic
uncertainties, we therefore plan to extend the current set of ensembles. We are
certain the configurations presented here will prove useful,  and we  are excited
about the interesting physics results that will be obtained.

\acknowledgments{
It is a pleasure to thank G.~Bali, B.~Leder, S.~Lottini, M.~L\"uscher, A.~Sch\"afer, R.~Sommer,
A.~Vladikas and H.~Wittig for many essential discussions and their help
in organizing this project. We are grateful to M.~L\"uscher and R.~Sommer for very
useful comments on earlier versions of this text.

We acknowledge PRACE for awarding us access to resource FERMI based in Italy at
CINECA, Bologna and to resource SuperMUC based in Germany at LRZ, Munich.
Furthermore, this work was supported by a grant from the Swiss National
Supercomputing Centre (CSCS) under project ID s384. We  are grateful for the
support received by the computer centers.

The D200 lattice has been produced on JUQUEEN.
The authors gratefully acknowledge the Gauss Centre for Supercomputing (GCS)
for providing computing time through the John von Neumann Institute for
Computing (NIC) on the GCS share of the supercomputer JUQUEEN at J\"ulich
Supercomputing Centre (JSC). GCS is the alliance of the three national
supercomputing centres HLRS (Universit\"at Stuttgart), JSC (Forschungszentrum
J\"ulich), and LRZ (Bayerische Akademie der Wissenschaften), funded by the
German Federal Ministry of Education and Research (BMBF) and the German State
Ministries for Research of Baden-W\"urttemberg (MWK), Bayern (StMWFK) and
Nordrhein-Westfalen (MIWF).

The ensembles C101r014 and C101r015 have been generated on the "Clover" HPC
Cluster at the Helmholtz Institute Mainz, University of Mainz. The runs at
$\beta=3.3$ have been performed on the PAX installation at DESY and the
iDataCool at Regensburg University. 

M.B., P.K., T.K. and S.S. are supported by the Deutsche Forschungsgemeinschaft (DFG)
in the SFB/TR~09 ``Computational Particle Physics''. 
G.P.E. acknowledges partial support by the MIUR-PRIN contract 20093BMNNPR and
G.H. acknowledges support by the the Spanish MINECO through the Ram\'on y
Cajal Programme and through the project FPA2012-31686 and by the Centro de
excelencia Severo Ochoa Program SEV-2012-0249. 
G.H. and H.H. acknowledge the support from the DFG in the SFB 1044.
M.P. acknowledges partial
support by the MIUR-PRIN contract 2010YJ2NYW and by the INFN SUMA project.
E.E.S, J.S., and W.S. are supported by the SFB/TRR-55 ``Hadron
Physics from Lattice QCD'' by the DFG.
E.E.S. also acknowledges support from the EU grant
PIRG07-GA-2010-268367.
}

\bibliography{cls}

\end{document}